\documentclass[aps,prd,showpacs,showkeys,preprintnumbers,twocolumn]{revtex4-1}

\usepackage{amsmath}
\usepackage{amssymb}
\usepackage{multirow}
\usepackage{graphicx}

\DeclareMathOperator{\sign}{sign}
\DeclareMathOperator{\diag}{diag}
\DeclareMathOperator{\tr}{Tr}

\begin{document}

\title{Nearest-Neighbour-Interactions from a minimal discrete flavour
  symmetry within $\mathsf{SU(5)}$ Grand Unification}

\date{\today}

\author{D.~Emmanuel-Costa}
\email{david.costa@ist.utl.pt}
\author{C.~Sim\~oes}
\email{csimoes@cftp.ist.utl.pt}

\affiliation{Departamento de F\'{\i}sica and Centro de F\'{\i}sica
  Te\'orica de Part\'{\i}culas, Instituto Superior T\'ecnico,
  Universidade T\'ecnica de Lisboa, Av. Rovisco Pais, 1049-001 Lisboa,
  Portugal}

\keywords{Flavour symmetries, Quark and lepton masses and mixing,
  Unified theories and models of strong and electroweak interactions,
  Neutrino mass and mixing}

\pacs{11.30.Hv, 12.15.Ff, 12.10.Dm, 14.60.Pq}

\preprint{arXiv:1102.3729, CFTP/11-004}

\begin{abstract}
  A flavour symmetry based on $\mathsf{Z}_4$ is analysed in the
  context of $\mathsf{SU(5)}$ Grand Unification with the standard
  fermionic content plus three right-handed neutrinos. The role of
  $\mathsf{Z}_4$ is to forbid some Yukawa couplings of up- and
  down-quarks to Higgs scalars such that the quark mass matrices
  $M_u,\,M_d$ have Nearest-Neighbour-Interaction~(NNI) structure, once
  they are generated through the electroweak symmetry breaking. It
  turns out in this framework that $\mathsf{Z}_4$ is indeed the
  minimal discrete symmetry and its implementation requires the
  introduction of at least two Higgs quintets, which leads to a two
  Higgs doublet model at low energy scale.
  Due to the $\mathsf{SU(5)}$ unification, it is shown that the
  charged lepton mass matrix develops also NNI form. However, the
  effective neutrino mass matrix exhibits a non parallel pattern, in
  the framework of the type-I seesaw mechanism. Analysing all possible
  zero textures allowed by gauge-horizontal symmetry
  $\mathsf{SU(5)}\times\mathsf{Z}_4$, it is seen that only two
  patterns are in agreement with the leptonic experimental data and
  they could be further distinguished by the light neutrino mass
  spectrum hierarchy.
  It is also demonstrated that $\mathsf{Z}_4$ freezes out the
  possibility of proton decay through exchange of colour Higgs triplets
  at tree-level.
\end{abstract}

\maketitle

\section{Introduction}
\label{sec:intro}

Grand Unified Theories~(GUTs) are beautiful attempts beyond the
Standard Model (SM) for understanding the observed quark and lepton
masses and their mixings, the so-called ``flavour puzzle''. This is
indeed corroborated by the fact that the running gauge couplings when
evolved to very large energy scales, typically $10^{14\,-\,16}$ GeV,
seem to unify to a unique coupling. The simplest GUT model which
accommodates the SM gauge and fermion fields in a few multiplets is
based on the group $\mathsf{SU(5)}$, proposed in 1974 by Georgi and
Glashow~\cite{Georgi:1974sy}. Some small extensions of the
Georgi-Glashow model are still viable
today~\cite{Dorsner:2005fq,Dorsner:2006hw,Dorsner:2006fx,Bajc:2006ia,
  Perez:2007rm}, even in its supersymmetric
version~\cite{Bajc:2002bv,Bajc:2002pg,EmmanuelCosta:2003pu,Perez:2007iw}.

Generically in GUT models, not only the SM gauge couplings do unify
but also the SM fermions are unified in a small number of large
multiplets that lead to new phenomenological signatures. An important
signature of most GUTs is the prediction for proton
decay~\cite{Nath:2006ut}, which has not yet been observed and then
severely constrains the GUT models. The fact that the quarks and
leptons are tied together in GUT multiplets is not enough to fully
determine the properties of their observed masses and
mixings. However, the GUT relations among quark and lepton Yukawa
matrices are an excellent starting point for building a flavour
symmetry. Thus, if one requires a flavour symmetry to enforce a
particular pattern in the up- and down-quark Yukawa couplings, this
engenders physical consequences in the leptonic sector.

During the last decades, a huge number of flavour symmetries with
different purposes have been extensively presented in the literature
(in the context of SM, Grand Unification, etc.). The simplest and
attractive possibility is to assume the vanishing of some Yukawa
matrix elements (``texture zeroes'') by the requirement of a discrete
symmetry~\cite{Branco:1987tv,Babu:2004tn,Grimus:2004hf,Low:2005yc,
  Ferreira:2010ir,Canales:2011ug}, such that it would naturally lead
to the flavour mixing angles be expressed in terms of mass ratios. The
converse is not necessarily true, since one can obtain zeroes in the
Yukawa
matrices~\cite{Branco:1999nb,Branco:2007nn,EmmanuelCosta:2009bx} just
by performing some set of transformations (weak basis transformations)
leaving the gauge sector diagonal. For instance, it is remarkable in
the SM that one can always go to a weak basis where both up- and
down-quark mass matrices $M_u,\,M_d$ have simultaneously the form or a
``parallel structure'',
\begin{equation}
\label{eq:nni}
M_{u,d}=\begin{pmatrix}
    0 & A_{u,d} & 0\\
    A_{u,d}^{\prime} & 0 & B_{u,d}\\
    0 & B_{u,d}^{\prime} & C_{u,d}
   \end{pmatrix}\,,
\end{equation}
known as the Nearest-Neighbour-Interaction~(NNI)
basis~\cite{Branco:1988iq}.  Being the matrix form in
Eq.~\eqref{eq:nni} for both quark sectors just a weak basis, no zero
in the NNI matrices has physical meaning. The NNI basis is closely
connected to the Fritzsch
ansatz~\cite{Fritzsch:1977vd,Li:1979zj,Fritzsch:1979zq}, which further
assumes the Hermiticity condition on the NNI quark mass matrices
$M_u,\,M_d$. Through a simple choice of weak basis transformation it
is always possible to make $M_u,\,M_d$ Hermitian, but their structures
are no longer of the NNI form. Assuming that $M_u,\,M_d$ are in the
NNI basis, it has been shown in Ref.~\cite{Branco:2010tx} that the
experimental data are still in agreement with relatively small
deviations from Hermiticity, at the 20\%~level.  This procedure was
also extended to the leptonic sector in Ref.~\cite{Fritzsch:2011cu}.

Furthermore, it was also shown in Ref.~\cite{Branco:2010tx} that it is
possible to attain the up- and down-quark mass matrices $M_u,\,M_d$
with NNI structure through the implementation of an Abelian discrete
flavour symmetry in the context of the two Higgs doublet
model~(2HDM). In that context, the minimal realisation is the group
$\mathsf{Z}_4$. In a general 2HDM, a NNI form for each Yukawa coupling
matrices cannot be a weak basis choice. Indeed, the requirement of the
$\mathsf{Z}_4$-symmetry does imply restrictions on the scalar
couplings to the quarks, although one gets no impact on the quark
masses and the Cabibbo-Kobayashi-Maskawa (CKM)
matrix~\cite{Cabibbo:1963yz,Kobayashi:1973fv}.

The purpose of this article is to study whether it is possible to
construct a $\mathsf{Z}_4$ flavour symmetry, similar to the one
implemented in Ref.~\cite{Branco:2010tx}, that leads to quark mass
matrices $M_u,\,M_d$ in the NNI form in the context of
$\mathsf{SU(5)}$ Grand Unification with the usual fermionic
content. In addition, since $\mathsf{SU(5)}$ implies relations among
quarks and leptons, we also explore the physical consequences of the
$\mathsf{Z}_4$ flavour symmetry on the leptonic sector for the case
where three right-handed neutrinos are added, being the type-I
seesaw~\cite{Minkowski:1977sc,Yanagida:1979as,GellMann:1980vs,
  Mohapatra:1979ia} the mechanism responsible for the light neutrinos
to acquire Majorana masses.

This article is organised as follows. In Section~\ref{sec:model} we
introduce the $\mathsf{SU(5)}\times\mathsf{Z}_4$ model. Next, in
Section~\ref{sec:proton} we analyse the different channels of proton
decay as well as the issues of unification in this model. Then in
Section~\ref{sec:quarklepton} we discuss in detail the form of the
leptonic mass matrix textures provided by the flavour symmetry. In
Section~\ref{sec:neutrino} we present our numerical analysis on the
leptonic sector as the result of the restrictions due to
$\mathsf{Z}_4$. The zero textures obtained for the effective neutrino
mass matrix are then confronted with leptonic observable
data. Finally, our conclusions are drawn in
Section~\ref{sec:conclusions}.


\section{The model}
\label{sec:model}

Following Ref.~\cite{Branco:2010tx}, we build an Abelian discrete
flavour symmetry within a $\mathsf{SU(5)}$ GUT model which yields at
low energies to a NNI form for both up- and down-quark mass
matrices. We choose the flavour symmetry to be Abelian, because it is
the simplest way to forbid some Yukawa couplings, so that texture
zeroes appear naturally in the mass matrices. The flavour symmetry is
also chosen to be discrete in order to avoid the presence of
Nambu-Goldstone bosons. To simplify our search for a minimal flavour
symmetry realisation on the full Lagrangian, we consider only the case
where the flavour group belongs to the $\mathsf{Z}_n$ family. Thus,
each fermionic or Higgs multiplet, $R$, transforms as
\begin{equation}
R\longrightarrow\,R^{\prime}=
e^{i\,\frac{2\pi}{n}\mathcal{Q}(R)}\,R\,,
\end{equation}
where the charges $\mathcal{Q}(R)\in\mathsf{Z}_n$.

The particle content of our GUT model is a small extension of the
original $\mathsf{SU(5)}$ model proposed by Georgi and
Glashow~\cite{Georgi:1974sy} in 1974. It contains three generations of
$\mathsf{10}$, $\mathsf{5}^{\ast}$ fermionic multiplets, which
accommodate the left-handed fermions of the SM,
$Q_i,u^c_i,d^c_i,L_i,e^c_i$, as follows
\begin{equation}
\label{eq:rep}
\mathsf{10}_i=(Q_i,u^c_i,e^c_i)\,,\quad
\mathsf{5}^{\ast}_i=(L_i,d^c_i)\,,
\end{equation}
where $i=1,2,3$ stands for the generation index. Furthermore, we
introduce three right-handed neutrinos $\nu^c_i$ (in the left-handed
picture), singlets under $\mathsf{SU(5)}$, as the simplest way to
generate the light neutrino masses needed to explain the observed
neutrino oscillation data. Being not constrained by any gauge
symmetry, the singlets $\nu^c_i$ can have a Majorana mass term and a
Dirac Yukawa term mixing with the leptonic doublets. After the
electroweak symmetry breaking, if one takes the Majorana mass term
close to the GUT scale, so much larger than the Dirac mass term, light
neutrino masses can be generated - the type-I seesaw
mechanism~\cite{Minkowski:1977sc,Yanagida:1979as,GellMann:1980vs,
  Mohapatra:1979ia}.

The Higgs sector of the model consists of an adjoint Higgs multiplet,
$\Sigma(\mathsf{24})$, chargeless under $\mathsf{Z}_n$, and two
quintets $H_1(\mathsf{5})$, $H_2(\mathsf{5})$, with different
$\mathsf{Z}_n$ charges $\phi_1$, $\phi_2$, respectively. Under these
charge assignments the full scalar potential $V$ reads as
\begin{equation}
\label{eq:pot}
\begin{split}
V&=
-\frac{1}{2}\mu^2\tr(\Sigma^2)+\frac{1}{3}a\tr(\Sigma^3)\\
&+\frac{1}{2}b^2\left[\tr(\Sigma^2)\right]^2+
\frac{\lambda}{4}\tr(\Sigma^4)\\
&+H_1^{\dagger}\left(\frac{1}{2}\mu_1^2+a_1\Sigma+
\lambda_{11}\tr(\Sigma^2)+\lambda_{12}\Sigma^2\right)H_1\\
&+H_2^{\dagger}\left(\frac{1}{2}\mu_2^2+a_2\Sigma+\lambda_{21}
\tr(\Sigma^2)+\lambda_{22}\Sigma^2\right)H_2\\
&+\lambda_{1}|H_1|^4+\lambda_{2}|H_2|^4+\lambda_3|H_1|^2|H_2|^2\\
&+\lambda_4\left(H_1^{\dagger}H_2H_2^{\dagger}H_1\right)\,,
\end{split}
\end{equation}
where the parameters $\mu$, $a$ and $b$ have mass dimensions, while
$\lambda$ is dimensionless. Note that the self-potential terms for the adjoint
field $\Sigma$ is as general as in the minimal $\mathsf{SU}(5)$ model.

The adjoint
field $\Sigma$ breaks spontaneously the
$\mathsf{SU(5)}$ gauge group to the SM group
($\mathsf{SU(3)_{\text{\sc c}} \times SU(2)_{\text{\sc l}} \times
  U(1)_{\text{\sc y}}}$), through the vacuum expectation value (VEV),
\begin{equation}
\label{eq:vevSM}
\langle\Sigma\rangle=\sigma\diag(2,2,2,-3,-3)\,,
\end{equation}
provided that $\sigma$ is the following solution
that minimises the scalar
potential~\cite{Li:1973mq,Buccella:1979sk,Guth:1979bh,Ruegg:1980gf}
given in Eq.~\eqref{eq:pot},
\begin{equation}
\label{eq:sigma}
\sigma= \frac{a}{2\lambda}\frac{1 \,+\, \sqrt{1 \,+\,
4\,\xi\,(60\eta\,+\,7)}} { 60\eta\, +\,7}\,,
\end{equation}
where $\eta\equiv b^2/\lambda$ and $\xi\equiv\lambda\mu^2/a^2\,$
requiring that $\eta>-7/60$ and $\lambda>0$. The parameters $\eta$ and
$\xi$ are enough to allow whether the VEV
from Eq.~\eqref{eq:vevSM} corresponds to an absolute minimum of the
potential. Thus, the parameters of the potential have to be properly chosen to
guarantee the SM group in the broken phase~\cite{Guth:1979bh} with the natural
value for $\sigma$ lying around the unification scale $\Lambda$. The
VEV given in Eq.~\eqref{eq:vevSM} also splits the adjoint Higgs field
$\Sigma$ in its components  $\Sigma_3$ (weak isospin triplet), $\Sigma_8$ (colour
octet) and $\Sigma_{24}$ (singlet), which become massive.

The Higgs quintets $H_{1},\,H_{2}$ are introduced to break the SM
gauge group down to $\mathsf{SU(3)_{\text{\sc c}}\times
  U(1)_{\text{e.m.}}}$ and also generate the fermion masses via the
Yukawa interactions at the electroweak scale. The quintets
$H_{1},\,H_{2}$ are split by the VEV $\langle\Sigma\rangle$ into the
Higgs doublets $\Phi_{1},\,\Phi_{2}$ and the Higgs colour-triplets
$T_{1},\,T_{2}$, respectively. Thus, the SM group is then broken
through the VEVs $v_1,v_2$ of the respective Higgs doublets
$\Phi_1,\Phi_2$, verifying
\begin{equation}
  v^2\,\equiv\, \left|v_1\right|^2 + \left|v_2\right|^2 =
  \left(\sqrt{2}\,G_F\right)^{-1}= (246.2\,\text{GeV})^2\,,
\end{equation}
where $G_F$ is the Fermi constant.

Furthermore, one has to avoid rapid proton decay mediated by the Higgs
colour-triplets $T_{1},\,T_{2}$, which can be solved by fine-tuning
the parameters of the Higgs potential,
$\mathcal{O}\left(\tfrac{v}{\sigma}\right)\sim10^{-(12\div 13)}$ - the
so-called doublet-triplet splitting problem. This fine-tuning can be
re-expressed as new constraints on the mass parameters $\mu_1^2$ and
$\mu_2^2$,
\begin{subequations}
\begin{align}
\mu_1^2=6\sigma\left( a_1  - 10 \sigma \lambda_{11} - 3 \sigma
\lambda_{12}\right)\,,\\
\mu_2^2=6 \sigma\left( a_2 - 10 \sigma \lambda_{21} - 3 \sigma
\lambda_{22}\right)\,,
\end{align}
\end{subequations}
such that the Higgs colour triplets $T_{1},\,T_{2}$ have masses:
\begin{subequations}
\begin{align}
m^2_{T_1}&=5\sigma(a_1-\sigma\lambda_{12})\,,\\
m^2_{T_2}&=5\sigma(a_2-\sigma\lambda_{22})\,.
\end{align}
\end{subequations}
Thus, once the heavy Higgs colour-triplets $T_{1},\,T_{2}$ are
integrated out, the model is just a two Higgs doublet model
(2HDM). Since the adjoint Higgs multiplet carries no $\mathsf{Z}_n$
charge, the obtained 2HDM automatically preserves the flavour symmetry
in higher orders of perturbation theory, provided that no
Nambu-Goldstone boson appears at tree-level due to an accidental
global symmetry~\cite{Georgi:1974au}.

In order to fully determine the $\mathsf{Z}_n$ charges for the
fermions and the Higgs quintets such that the quark mass matrices
$M_u,\,M_d$ have NNI form, one needs to analyse the Yukawa
interactions~\cite{Branco:2010tx}. The most general Yukawa Lagrangian
reads as
\begin{equation}
\label{eq:yukawa}
\begin{split}
-\mathcal{L}_\text{Y} &=
\frac{1}{4}\left(\Gamma^1_u\right)_{ij}10_i\,10_j\,H_1
+\frac{1}{4}\left(\Gamma^2_u\right)_{ij}10_i\,10_j\,H_2\\
& +
\sqrt2\left(\Gamma^1_d\right)_{ij}10_i\,5^{\ast}_j\,H_1^{\ast} +
\sqrt2\left(\Gamma^2_d\right)_{ij}10_i\,5^{\ast}_j\,H_2^{\ast}\\
& +
\left(\Gamma^1_D\right)_{ij}\,5^{\ast}_i\,\nu^c_j\,H_1+
\left(\Gamma^2_D\right)_{ij}\,5^{\ast}_i\,\nu^c_j\,H_2\\
& +\frac{1}{2}\left(M_R\right)_{ij}\nu^c_i\,\nu^c_j+\text{H.c.}\,,
\end{split}
\end{equation}
where $\Gamma^{1,2}_u$ and $M_R$ are symmetric complex matrices, while
$\Gamma^{1,2}_{d,\:D}$ are just general complex matrices.  The quark
mass matrices $M_u,\,M_d$ are then given by
\begin{subequations}
\label{eq:masses}
\begin{align}
\label{eq:mu}
 M_u =& v_1\,\Gamma^1_u+v_2\,\Gamma^2_u\,,\\
 M_d =& v_1^{\ast}\,\Gamma^1_d+v_2^{\ast}\,\Gamma^2_d\,,
\end{align}
\end{subequations}
and their zeroes are directly settled by the Yukawa matrices
$\Gamma^{1,2}_{u,d}\,$. It is clear from Eq.~\eqref{eq:mu} that 
the mass matrix $M_u$ is symmetric.

In order to guarantee that the $(33)$-element of up-quark matrix $M_u$
does not vanish we set the charge $\phi_2$ as
\begin{equation}
\label{eq:constr}
 \phi_2\,=\,-2q_3\,,
\end{equation}
where $q_3\equiv\mathcal{Q}(\mathsf{10}_3)$. This choice automatically
fixes the $\mathsf{Z}_n$ charges of the multiplets
$\mathsf{5}^{\ast}_i,\, \mathsf{10}_i$ as a function of $\phi_1$ and
$q_3$ as
\begin{equation}
\begin{aligned}
\label{eq:su5q}
\mathcal{Q}(\mathsf{10}_i)&=(3q_3+\phi_1,\,-q_3-\phi_1,\,q_3)\,,\\
\mathcal{Q}(\mathsf{5}^{\ast}_i)&=(q_3+2\phi_1,\,-3q_3,\,-q_3+\phi_1)\,.
\end{aligned}
\end{equation}
The $\mathsf{Z}_n$ charges of the right-handed neutrinos, $\nu_i^c$
(singlets under $\mathsf{SU(5)}$) have no restrictions and are taken
as free parameters,
\begin{equation}
\label{eq:chnu}
\mathcal{Q}(\nu^c_i)=(\nu_1,\nu_2,\nu_3)\,.
\end{equation}
It is in fact not enough to have the charge assignments given in
Eq.~\eqref{eq:su5q} in order to guarantee the NNI structure for the
mass matrices $M_u,\,M_d$, one has in addition to preserve the
NNI-zero entries, which implies that one should forbid the quark
bilinears corresponding to the NNI-zero entries to couple to Higgs
doublets. The analysis of the quark bilinears can be made directly
from the $\mathsf{10}_i\mathsf{10}_j$ and
$\mathsf{10}_i\mathsf{5}^{\ast}_j$ bilinears. By taking into account
Eq.~(\ref{eq:rep}) one can derive the $\mathsf{10}_i\mathsf{10}_j$
bilinear charges as
\begin{equation}
\label{eq:biluc}
\begin{pmatrix}
6q_3+2\phi_1 & 2q_3 & 4q_3+\phi_1\\
2q_3 & -2\phi_1-2q_3 & -\phi_1\\
4q_3+\phi_1 & -\phi_1 & 2q_3
\end{pmatrix}
\end{equation}
and the $\mathsf{10}_i\mathsf{5}^{\ast}_j$ bilinear charges as
\begin{equation}
\label{eq:bildc}
\begin{pmatrix}
4q_3+3\phi_1 & \phi_1 & 2\phi_1+2q_3\\
\phi_1 & -4q_3-\phi_1 & -2q_3\\
2\phi_1+2q_3 & -2q_3 & \phi_1
\end{pmatrix}\,.
\end{equation}
One sees immediately from Eq.~(\ref{eq:biluc}), that in the case of
$\phi_1=\phi_2$ one gets that both Higgs doublets $\Phi_1$,
$\Phi_2$ can couple to the bilinear $\mathsf{10}_1\mathsf{10}_1$,
thus destroying the NNI form for $M_u$. According to the
Eqs.~(\ref{eq:biluc}) and (\ref{eq:bildc}), the lowest order group of
$\mathsf{Z}_n$-type which respects the zero-entries in the NNI form is
$\mathsf{Z}_4$, as in Ref.~\cite{Branco:2010tx}. The symmetry group
$\mathsf{Z}_4$ is indeed the minimal discrete flavour symmetry that
gives rise to NNI structure for $M_u,\,M_d$, since the other
candidate, $\mathsf{Z}_2 \times \mathsf{Z}_2$, having all its
non-trivial elements of order~2, is not viable.

The zero-entries in the Yukawa coupling matrices $\Gamma^{1,2}_{u,d}$
are unequivocally determined by Eq.~\eqref {eq:su5q}, provided that
the bilinear entries given in Eqs.~\eqref {eq:biluc}
and~\eqref{eq:bildc} are taken correctly into account, as follows:
\begin{subequations}
\label{eq:Y}
\begin{align}
\Gamma^1_{u}&=
\begin{pmatrix}
0 & 0 & 0\\
0 & 0 & b_{u}\\
0 & {b}_{u}& 0
\end{pmatrix}\,,\quad
\Gamma^2_{u}=
\begin{pmatrix}
0 & a_{u} & 0\\
a_{u} & 0 &0\\
0 & 0 & c_{u}
\end{pmatrix}\,,\\[2mm]
\Gamma^1_{d}&=
\begin{pmatrix}
0 & a_{d} & 0\\
a^{\prime}_{d} & 0 &0\\
0 & 0 & c_{d}
\end{pmatrix}\,,\quad
\Gamma^2_{d}=
\begin{pmatrix}
0 & 0 & 0\\
0 & 0 & b_{d}\\
0 & b^{\prime}_{d}& 0
\end{pmatrix}\,,
\end{align}
\end{subequations}
which when applying Eqs.~\eqref{eq:masses} leads to mass matrices
$M_u,M_d$ with a NNI form. We point out that the NNI structures thus
obtained have a different nature than in the pure SM case. As we have
emphasised in the introduction, in the SM the NNI structure for $M_u$,
$M_d$ is just a choice of weak basis. Instead in our model, the NNI
form for $M_u$, $M_d$ arises as the requirement of a $\mathsf{Z}_4$
symmetry, i.e. the form of the Yukawa couplings $\Gamma_{u,d}^{1,2}\,$
are dictated by $\mathsf{Z}_4$ in Eqs.~\eqref{eq:Y}, and therefore it
is not a weak basis choice. In fact, the $\mathsf{Z}_4$ do imply new
restrictions on the scalar couplings to quarks.

In order to complete the construction of the flavour symmetry
$\mathsf{Z}_4$, one has to ensure in addition that the scalar
potential does not acquire an extra accidental global symmetry which,
upon spontaneous electroweak symmetry breaking, would lead to a
massless Nambu-Goldstone boson at tree
level~\cite{Weinberg:1972fn}. It is indeed true that the most general
scalar potential invariant under $\mathsf{SU}(5)\times\mathsf{Z}_4$,
given in Eq.~\eqref{eq:pot}, possesses an extra $\mathsf{U}(1)$ global
invariant transformation on the quintet fields. Analogously of what
was described in~\cite{Branco:2010tx}, this problem can be cured by
soft-breaking the $\mathsf{Z}_4$ symmetry through the introduction of
a term like
\begin{equation}
\label{eq:soft}
\mu_{12}^2\, H^{\dagger}_1\,H_2\, +\text{H.c.}\,,
\end{equation}
in the scalar potential in Eq.~\eqref{eq:pot}. Notice that the
soft-breaking term $\mu_{12}$ is not sufficient to break CP spontaneously
in this model, so that CP violation arises simply from complex Yukawa
couplings, leading to the Kobayashi-Maskawa mechanism, see KM in
Ref.~\cite{Kobayashi:1973fv}.

An elegant way to avoid the term given in Eq.~\eqref{eq:soft} and at
the same time prevent the existence of the global $\mathsf{U}(1)$
symmetry is simply by adding a complex $\mathsf{SU}(5)$ singlet Higgs
field $S$ which transforms non-trivially under $\mathsf{Z}_4$. The
most general potential involving the singlet field $S$ reads as
\begin{equation}
\begin{split}
\label{eq:potS}
V_{S}&=\,\left[H^{\dagger}_1\left(\mu^{\prime}_{12}\,+\,\lambda^{\prime}_{
12}\Sigma\right)H_2\,S\,+\,H.c.\right]
-\frac{1}{2}\,\mu_S^2\,|S|^2\\
&\,+\,\lambda_S\,|S|^4\,+\,\lambda^{\prime}_S(S^4\,
+\, H.c.)\,.
\end{split}
\end{equation}
Note that the last term in the Eq.~\eqref{eq:potS}, allowed by
$\mathsf{Z}_4$, explicitly prevents the appearance of Goldstone bosons
that could result from an accidental global $\mathsf{U(1)}$. In the
case of the VEV of the complex singlet field $S$ is of the order of
the GUT scale, it is easy to verify that $\arg(\langle S\rangle)=0$ or
$\pi/4$, but it does not break the CP symmetry spontaneously.

We are now able to derive the texture zeroes in the leptonic
sector from the constructed $\mathsf{SU}(5)\times\mathsf{Z}_4$ model.
Since in this model the relation
\begin{equation}
\label{eq:mdml} M_e=M_d^{\top}\,,
\end{equation}
holds at the GUT scale and the zeroes of the NNI structure are placed
symmetrically, one easily sees that the flavour symmetry
$\mathsf{Z}_4$ also leads to a charged lepton mass matrix $M_{e}$ with
NNI form like for quark mass matrices $M_u,\,M_d$. On the other hand,
for the neutrino sector we do not expect a parallel structure neither
for the Dirac mass matrix, $m_D$,
\begin{equation}
  m_D = v_1\,\Gamma^1_D+v_2\,\Gamma^2_D\,,
\end{equation}
nor for the Majorana mass matrix $M_R$, since the $\mathsf{Z}_4$
charges of the right-handed neutrino are taken arbitrary.

It is well-known that the GUT relation given by Eq.~\eqref{eq:mdml} is
not in fact compatible with the down-type quark and the charged lepton
mass hierarchies observed at low energies (badly violated for the
first and second generations)~\cite{Buras:1977yy}. In the following
subsections, we shall roughly sketch two small extensions of the model
that have the aim to modify properly the GUT relation given in
Eq.~\eqref{eq:mdml} without changing the zeroes in the quark mass
matrices $M_u,M_d$ and the charged lepton mass matrix $M_e$.

\subsection{``Consistent'' $\mathsf{SU}(5)$}
\label{sub:consistent}

A possible way to modify correctly the relation given by
Eq.~\eqref{eq:mdml} without adding new representations is by
considering some non-renormalisable higher dimensional
operators~\cite{Bajc:2002bv,EmmanuelCosta:2003pu},
$\mathcal{O}(1/\Lambda')$, due to physics above the GUT scale,
$\Lambda'\gg\Lambda$. The natural scale for $\Lambda'$ is to be two or
three orders of magnitude greater than the GUT scale. Higher
dimensional operators involving the adjoint field $\Sigma$ of the type
\begin{equation}
  \label{yukawa-planck}
\sum_{n=1,2}\frac{\sqrt{2}}{\Lambda'}
      \left(\Delta_n\right)_{ij}\, H^{\ast}_{n\,a}\, \mathsf{10}_i^{ab}\,
      \Sigma_b^c\, \mathsf{5}^{\ast}_{jc}\,,
\end{equation}
with the indices $a,b,c=1,\cdots,5$ and $i,j=1,2,3$, would
contribute to the relation in Eq.~\eqref{eq:mdml} as
\begin{equation}
  \label{sm-planck}
  M_d-M_{e}^{\top} = 5\frac{\sigma}{\Lambda'}
(v_1^{\ast}\,\Delta_1+v_2^{\ast}\,\Delta_2)\,.
 \end{equation}
 The complex matrices $\Delta_1$ and $\Delta_2$ can account for the
 discrepancies between $M_d$ and $M_{e}$. The quark and charged lepton
 mass matrices $M_u,M_d,M_e$ remain in the NNI form, since the adjoint
 field $\Sigma$ is trivial under $\mathsf{Z}_4$ and contributes to the
 quark Yukawa matrices $\Gamma^{1,2}_{u,d}$ through dimension-five
 operators when it acquires VEV.  The only higher dimensional
 operators that could spoil the NNI structure on the quark mass
 matrices are of dimension-six, e.g.
\begin{equation}
\frac{\lambda}{{\Lambda'}^2}\, 10_2\,10_2\,H_1\,H^{\ast}_1\,H_2\,,
\end{equation}
 and therefore very much suppressed,
$\mathcal{O}(v^2/{\Lambda'}^2)\,$. Moreover, the presence of
dimension-five operators in the Higgs potential given in
Eq.~\eqref{eq:pot} can contribute to the splitting between the masses
of the Higgs multiplets $\Sigma_3$ and $\Sigma_8$ by several orders of
magnitude.

\subsection{Adjoint  $\mathsf{SU}(5)$}
\label{sub:adjoint}

The other alternative consists in maintaining the full Lagrangian
renormalisable just by requiring a $\mathsf{45}$ dimensional Higgs
scalar~\cite{Perez:2007rm}, $\mathcal{H}(\mathsf{45})$, instead of the
quintet $H_2$.  The field representation
$\mathcal{H}^{\alpha\,\beta}_{\gamma}$ satisfies the relations:
$\mathcal{H}^{\alpha\,\beta}_{\gamma}=-\mathcal{H}^{\beta\,\alpha}_{
  \gamma}$ and
$\sum^5_{\alpha=1}{\mathcal{H}^{\alpha\,\beta}_{\alpha}=0}\,$.  Thus,
the potential given in Eq.~\eqref{eq:pot} is now modified by the
following terms:
\begin{equation}
\begin{split}
&\mathcal{H}^{\dagger}\left(\frac{1}{2}\mu_{\mathcal{H}}^2+\lambda_{21}
\tr(\Sigma^2)+a_2\Sigma+\lambda_{22}\Sigma^2\right)\mathcal{H}
\\
&+\lambda_{2}\,(\mathcal{H}^{ \dagger}\mathcal{H})^2 +
\lambda_3|H_1|^2(\mathcal{H}^{\dagger}\mathcal{H})\\
&+\lambda_4\left(H_1^{\dagger}\mathcal{H}\mathcal{H}^{\dagger}
H_1\right)\,,
\end{split}
\end{equation}
that substitute the terms involving the quintet $H_2$.  For the sake
of simplicity of the notation, the $\mathsf{SU}(5)$ invariant
contractions involving $\mathcal{H}$ were not explicitly written. The
mismatch between $M_d$ and $M_{e}$ is explained as
\begin{equation}
M_d-M_{e}^{\top}=8\,\Gamma^2_d\,v^{\ast}_{45}\,,
\end{equation}
where $v_{45}$ is the strength of the vacuum expectation value of the
field $\mathcal{H}$, assuming
\begin{equation}
 \langle\mathcal{H}^{\beta\,5}_{\alpha}\rangle
=v_{45}\left(\delta_{\alpha}^{\beta}-4\,\delta_{4}^{\alpha}
\delta_{\beta}^{4}\right)\,.
\end{equation}
It is clear that if the Higgs multiplet $\mathcal{H}$ has the same
$\mathsf{Z}_4$ charge as the one assigned for $H_2$ in
Eq.~\eqref{eq:constr}, one recovers the NNI form for the quark mass
matrices $M_u,\,M_d$ and the charged lepton mass matrix $M_e$.


\section{Proton decay and Unification}
\label{sec:proton}

In this section we analyse the proton decay in the constructed
$\mathsf{SU}(5) \times \mathsf{Z}_4$ model. In this model, the proton
decays through the exchange of the heavy lepto-quark gauge bosons
$X,Y$ or the colour Higgs triplets $T_1,T_2$. The experimental limits
on the proton decay rate severely constrain the masses of such heavy
states that we shall assume of the order of the unification scale
$\Lambda$, since in this scenario the proton decay rate is inversely
proportional to the mass square of the heavy states. On the other
hand, the unification scale is by definition the scale where the
running gauge couplings measured at the scale
$M_Z=91.1876\pm0.0021\,\text{GeV}$~\cite{Nakamura:2010zzi} do
unify. Thus, the limits on the proton decay rate have to be confronted
with the parameters that govern the evolution of the gauge couplings.

The twelve lepto-quark gauge bosons $X,Y$ (components of a colour weak
isospin doublet) arise from the adjoint $\mathsf{24}$ representation
that also contains the twelve gauge bosons of the SM. The gauge bosons
$X,Y$ become massive through the Higgs mechanism with a common mass,
$M_V$,
\begin{equation}
M_V=\frac{25}{8}g_U^2\sigma^2\,,
\end{equation}
where $g_U$ is the unified gauge coupling. To suppress the $X,Y$ boson
proton decay channels, one has necessarily that $M_V\gg m_p$ (the
proton mass) which then leads to an approximate four fermion
interaction (dimension-six operators) proportional to $1/M^2_V$ and
the unified coupling $\alpha_U\equiv g_U^2/4\pi\,$.  In this
approximation, the proton decay width can be estimated
as~\cite{Langacker:1980js}:
\begin{equation}
\Gamma\approx \alpha_U^2 \frac{m_p^5}{M_V^4}\,.
\end{equation}
Making use of the most restrictive constraints on the partial proton
lifetime $\tau(p\rightarrow \pi^0 e^+)>8.2\times10^{33}$
years~\cite{Nakamura:2010zzi}, one can derive a rough lower bound for
the $X,Y$ mass scale $M_V$,
\begin{equation}
\label{eq:lower}
M_V>(4.0-5.1)\times10^{15}\,\text{GeV}\,,
\end{equation}
which corresponds a range of the unified gauge coupling
$\alpha_U^{-1}\approx 25 - 40$, as suggested by performing the
renormalisation group evolution of the gauge couplings (see details in
Appendix~\ref{a:beta}). Since we assume for the unification scale
$\Lambda\sim M_V$, the constraint given by Eq.~\eqref{eq:lower}
determines the scale where the gauge couplings should unify (for a
recent review see~\cite{Nath:2006ut}).

Usually in non-supersymmetric scenarios, the proton decay through the
exchange of Higgs colour triplets $T_1,T_2$ is very suppressed. Being
these decay modes also described by dimension-six operators, their
suppression is proportional to products of Yukawa couplings,which are much
smaller than the gauge couplings in the $X,Y$ boson exchange. In fact,
the contribution of these dimension-six operators vanishes at
tree-level when the $\mathsf{Z}_4$ symmetry is exact. The dimension-six
operators contributing to the proton decay via the colour triplet
exchange are given at tree-level by:
\begin{equation}
\label{eq:pdecay}
\sum_{n=1,2}
\frac{\left(\Gamma^n_u\right)_{ij}\left(\Gamma^n_d\right)_{kl}}{M^2_{T_n}}
\left[\frac{1}{2} (Q_iQ_j)(Q_kL_l)+(u^c_ie^c_j)(u^c_kd^c_l)\right]\,.
\end{equation}
It is then clear from the pattern of the Yukawa coupling matrices
$\Gamma^{1}_{u}$ and $\Gamma^{2}_{d}$ given in Eqs.~\eqref{eq:Y} that
the only possible non-vanishing contribution of the dimension-six
operators given in Eq.~\eqref{eq:pdecay} involve necessarily fermions
of the third generation. One concludes that at tree-level the proton
does not decay through the four-fermion interactions described by the
operators given in Eq.~\eqref{eq:pdecay}.

If one soft-breaks the $\mathsf{Z}_4$ symmetry of the potential with a $\mu_{12}$-term as the one
written in Eq.~\eqref{eq:soft}, such term mixes already at tree-level
the heavy states $T_1$, $T_2$ and therefore induces proton decay
through dimension-six operators proportional to
$\left(\Gamma^2_u\right)_{ij}\left(\Gamma^1_d\right)_{kl}$, thus
involving fermions of the first and the second generations. This can
be avoided when a singlet scalar $S$, charged under $\mathsf{Z}_4$, is
introduced in the potential as shown in Eq.~\eqref{eq:potS} and in
this case proton decay via $T_1,T_2$ exchange vanishes once more at
tree-level.

In what concerns unification, it is widely established that the
running gauge couplings do not unify in the context of the SM even in
the presence of an extra Higgs doublet~\cite{Nath:2006ut}. However, if
one considers the splitting between the masses of the multiplets
$\Sigma_3$ and $\Sigma_8$, it turns out possible to unify the gauge
couplings at two-loop level even without taking into account threshold
effects. We looked for unification assuming the masses of the bosons
$X$, $Y$, $T_1$ and $T_2$ at the unification scale $\Lambda$ and
allowing a splitting between the masses $M_{\Sigma_3}$,
$M_{\Sigma_8}$. We also assume the masses of the Higgs doublets
$\Phi_1,\Phi_2$ lying around the electroweak scale.

From our numerics, we have found a small variation for the unification
scale $\Lambda$,
\begin{equation}
\label{eq:uni}
 1.3\times 10^{14}\,\text{GeV}\leq \Lambda \leq 2.4\times
10^{14}\,\text{GeV}\,,
\end{equation}
and for the masses $M_{\Sigma_3}$ and
$M_{\Sigma_8}$,
\begin{subequations}
\begin{align}
  M_Z\leq M_{\Sigma_3}\leq 1.8\times 10^4\,\text{GeV}\,,\\
  5.4\times 10^{11}\,\text{GeV}\leq M_{\Sigma_8}\leq 1.3\times
10^{14}\,\text{GeV}\,,
\end{align}
\end{subequations}
which corresponds to a mass difference of the order
$M_{\Sigma_8}/M_{\Sigma_3}=\mathcal{O}(10^{9-10})\,$. These results do
not get improved even when a large splitting between the two Higgs
doublets is considered.  The details concerning the equations of gauge
coupling evolution are sketched in Appendix~\ref{a:beta}. For
illustration, in Fig.~\ref{fig1} we plot the gauge coupling evolution
at two-loop level for $M_{\Sigma_3}=500$ GeV, which fixes
$\Lambda=1.9\times 10^{14}$ GeV and $M_{\Sigma_8}=3.2\times 10^{12}$
GeV in order to achieve unification.

\begin{figure}[floatfix]
\begin{center}
\includegraphics[width=8.5cm,clip]{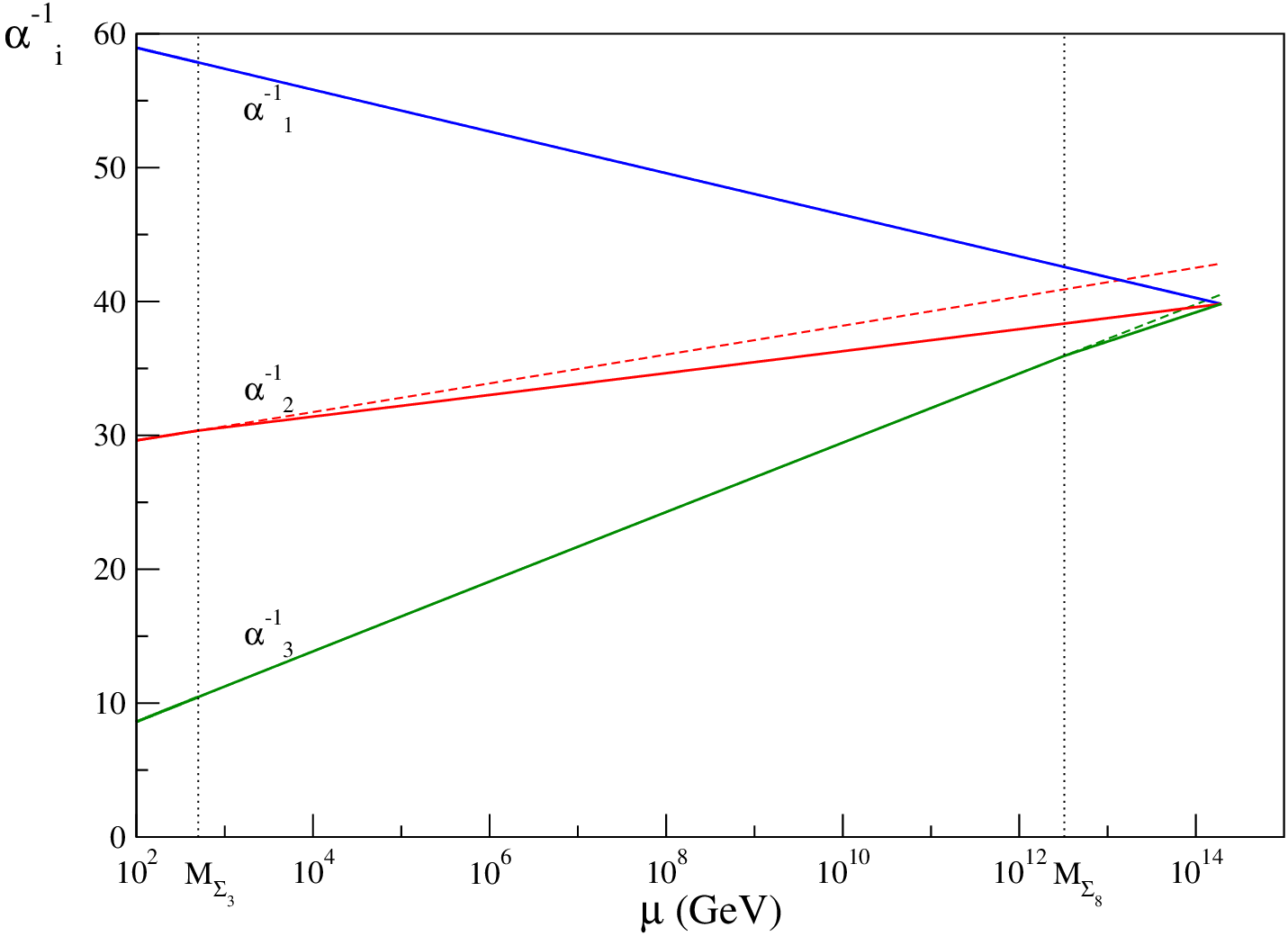}
\end{center}
\caption{\label{fig1} The plot of the gauge coupling evolution,
  $\alpha^{-1}_{1,2,3}$ , at two-loop level for the choice of
  $M_{\Sigma_3}=500$~GeV. The unification then occurs for
  $\Lambda=1.3\times 10^{14}$ GeV and $M_{\Sigma_8}=3.2\times
  10^{11}$~GeV. The dashed line corresponds to the running of the
  gauge couplings without considering the intermediate scales
  $M_{\Sigma_3}$ and $M_{\Sigma_8}$.}
\end{figure}

The unification of the gauge couplings suffers from two potential
problems. First, the scale $\Lambda$ in Eq.~\eqref{eq:uni} seems to
be lower than what is required by the lower bound given in
Eq.~\eqref{eq:lower}. Second, the mass splitting between
$M_{\Sigma_3}$ and $M_{\Sigma_8}$ is unnaturally large. This may
suggest the necessity of introducing extra multiplets in order to
relax such constraints, e.g. see
Refs.~\cite{Dorsner:2005fq,Perez:2007rm}. It is interesting to remark
that the alternative presented in subsection~\ref{sub:adjoint} can
successfully adjust the running of SM gauge couplings, since the
broken components of the $\mathsf{45}$ Higgs representations have the
correct quantum numbers~\cite{Perez:2007rm} for the unification of the
gauge couplings.


\section{Quark and lepton mass matrices}
\label{sec:quarklepton}

In Section~\ref{sec:model}, we have derived a $\mathsf{Z}_4$ flavour
symmetry that induces the quark mass matrices $M_u,\,M_d$ to have a
NNI structure in the framework of $\mathsf{SU(5)}$ Grand Unification
with minimal fermion content. Due to the fact that quarks and leptons
are tied together in fermionic multiplets, the flavour symmetry
imposed at the Lagrangian also restricts the form of the charged
lepton mass matrix $M_e$, which develops a NNI structure. Moreover,
the extensions considered in subsections \ref{sub:consistent} and
\ref{sub:adjoint} not only lead to $M_d\neq M_e^{\top}$ but also $M_u$
is no longer
symmetric~\cite{Langacker:1980js,Bajc:2002bv,EmmanuelCosta:2003pu}. Hence,
in what follows we shall assume arbitrary NNI mass matrices $M_u$,
$M_d$ and $M_e$, where the asymmetry among the elements $A_{u,d,e}$
and $A^{\prime}_{u,d,e}$, $B_{u,d,e}$ and $B^{\prime}_{u,d,e}$ can be
measured by the deviation parameters $\epsilon^{u,d,e}_{a,b}$ as
\begin{subequations}
\begin{align}
\label{eq:eaeb}
\epsilon^{u,d,e}_a&\equiv\frac{\left|A^{\prime}_{u,d,e}\right|-\left|A^{\,}_{u,d,e}\right|}{
  \left|A^{\prime}_{u,d,e}\right|+\left|A^{\,}_{u,d,e}\right|}\,,
\\\nonumber\\
\epsilon^{u,d,e}_b&\equiv\frac{\left|B^{\prime}_{u,d,e}\right|-\left|B^{\,}_{u,d,e}\right|}{
  \left|B^{\prime}_{u,d,e}\right|+\left|B^{\,}_{u,d,e}\right|}\,,
\end{align}
\end{subequations}
where the parameters $A_e$, $A^{\prime}_e$, $B_e$ and $B^{\prime}_e$
follow the convention made in Eq.~\eqref{eq:nni} for the quark sector. Global measurements
of the asymmetry in the quark, $\varepsilon_q$, and charged lepton,
$\varepsilon_e$, sectors are defined as
\begin{subequations}
\label{eq:eps}
\begin{align}
 \varepsilon_q&\equiv\frac12\sqrt{
(\epsilon^u_a)^2+(\epsilon^u_b)^2+(\epsilon^d_a)^2+(\epsilon^d_b)^2}\,,
\\
 \varepsilon_{e}&\equiv\sqrt{\frac{(\epsilon^{e}_a)^2
+(\epsilon^{e}_b)^2}2}\,,
\end{align}
\end{subequations}
It is interesting to see that in the limit when
$\varepsilon_q=\varepsilon_e=0$ one recovers the Fritzsch
ansatz~\cite{Fritzsch:1977vd,Li:1979zj,Fritzsch:1979zq}.

Concerning the neutrino sector, since neutrino charges are taken as
free parameters in Eq.~\eqref{eq:chnu}, one has to classify all viable
textures for the mass matrices $m_D$ and $M_R$ by scanning all
combinations of the $\mathsf{Z}_4$ neutrino charges. Hence, no NNI
form is expected for the mass matrices $m_D$ and $M_R$ and their
resulting textures have to be confronted with the neutrino
experimental data.  We start our scanning by deriving the allowed
range for the Higgs doublet charges $\phi_1$, $\phi_2$. As dictated by
Eq.~\eqref{eq:constr}, the charge of the Higgs doublet $\Phi_2$ can
only take two values: $\phi_2=0$ or $2\,$. On the other hand, the
charge of the Higgs doublet $\Phi_1$ has to be odd ($\phi_1=1$
or~$3\,$), which can be clearly seen, for instance, by noting that the
entries (2,2) and (3,3) of the bilinear given in Eq.~\eqref{eq:biluc}
would be equal when $\phi_1$ is even. Once the $\mathsf{Z}_4$ charges
of the right-handed neutrino fields, $\nu_i$, are fixed, one can
immediately determine the pattern of the effective neutrino mass matrix
$m_{\nu}$. From the structure of the Lagrangian in
Eq.~\eqref{eq:yukawa} one can derive the effective neutrino mass
matrix $m_{\nu}$, which is given by the usual standard type-I seesaw
formula~\cite{Minkowski:1977sc,Yanagida:1979as,GellMann:1980vs,
  Mohapatra:1979ia}:
\begin{equation}
\label{eq:seesaw}
m_{\nu}=-m_D\,M^{-1}_R\,m^{\top}_D\,,
\end{equation}
to an excellent approximation. Since the symmetric Majorana
mass matrix $M_R$ is directly introduced at the Lagrangian level,
its pattern is determined by the charges $\nu_i$. Thus the
$\mathsf{Z}_4$ charges of bilinears $\nu^c_i\,\nu^c_j$ are given by
\begin{equation}
 \label{eq:bilnucnuc}
\begin{pmatrix}
2 \nu_1 & \nu_1+\nu_2 & \nu_1+\nu_3\\
\nu_2+\nu_1 & 2 \nu_2 & \nu_2+\nu_3\\
\nu_3+\nu_1 & \nu_3+\nu_2 & 2 \nu_3
\end{pmatrix}\,.
\end{equation}
The texture zeroes in the Dirac neutrino mass matrix $m_D$ are then
obtained thereby computing the charges of the bilinears
$\mathsf{5}^{\ast}_i\,\nu^c_j\,$,
\begin{equation}
 \label{eq:bilnuc}
\begin{pmatrix}
q_3+2\phi_1+\nu_1 & q_3+2\phi_1+\nu_2 & q_3+2\phi_1+\nu_3\\
-3q_3+\nu_1 & -3q_3+\nu_2 & -3q_3+\nu_3\\
-q_3+\phi_1+\nu_1 & -q_3+\phi_1+\nu_2 & -q_3+\phi_1+\nu_3
\end{pmatrix}\,
\end{equation}
and verifying their couplings to the Higgs doublets. Finally, one is
then able to compute the texture zeroes in the effective neutrino mass
matrix, $m_{\nu}$, just by applying the seesaw formula given in
Eq.~\eqref{eq:seesaw}. Thus, by spanning all allowed values for the
charges $\phi_1$, $q_3$ and $\nu_i$, one can draw all possible zero
textures allowed by the symmetry $\mathsf{Z}_4$.  We sketched in
Table~\ref{tab:results} all possible effective neutrino mass matrices
$m_{\nu}$ obtained as a function of the parameters $\phi_1$, $q_3$ and
$\nu_i$. The notation used to represent all textures in
Table~\ref{tab:results} reflects the fact that each of them is related
to one of the following three classes of textures:
\begin{equation}
\label{eq:textures}
\left.\rm{I}\right)\,\begin{pmatrix}
    0 & A & 0\\
    A & 0 & B\\
    0 & B & C
   \end{pmatrix},
\:\:
\left.\rm{II}\right)\,\begin{pmatrix}
0 & A & 0\\
A & B & C \\
0 & C & D
 \end{pmatrix},
\:\:
\left.\rm{III}\right)\,\begin{pmatrix}
A & 0 & 0\\
0 & B & C\\
0 & C & D
\end{pmatrix},
\end{equation}
through a permutation matrix $P_g$ of the form
\begin{equation}
\label{eq:permut}
m^{\,\,_{(g)}}_{\nu}=P_{g}\,m_{\nu}\,P^{\top}_{g}\,,\quad g\in S_3\,,
\end{equation}
where $\{P_{g}\}$ are the six $3\times3$ permutation matrices
isomorphic to the symmetric group $S_3$. This identification is also
useful in simplifying the diagonalisation procedure. It is clear from
the seesaw formula given in Eq.~\eqref{eq:seesaw} that any permutation
among the right-handed neutrino $\mathsf{Z}_4$ charges does not change
the pattern of the effective neutrino mass matrix $m_{\nu}$. This is
indeed the reason why we have denoted the charges $\nu_i$ in
Table~\ref{tab:results} just by one ordered 3-tuple.

\begin{table}[floatfix]
\caption{\label{tab:results} The obtained textures zeroes for the effective
neutrino mass matrices in the context of $\mathsf{SU(5)} \times \mathsf{Z}_4$
symmetry with two-Higgs doublets.}
\begin{ruledtabular}
\begin{tabular}{ccccc}
& $q_3$ & $\nu=(0,1,3)$ & $\nu=(1,2,3)$ & $\nu_{i}\in\{0,2\}$ \\[1ex]
\hline\\[-2ex]
\multirow{4}{*}{\small $\phi_1=1$} & $0$ & $\rm{I}_{(132)}$ & $\rm{II}_{(12)}$ &
$\rm{III}_{(12)}$ \\[1ex]
& $1$ & $\rm{I}_{(13)}$  & $\rm{II}$        & $\rm{III}$\\[1ex]
& $2$ & $\rm{II}_{(12)}$ & $\rm{I}_{(132)}$ & $\rm{III}_{(12)}$\\[1ex]
& $3$ & $\rm{II}$        & $\rm{I}_{(13)}$  & $\rm{III}$\\[1ex]
\hline\\[-2ex]
\multirow{4}{*}{\small $\phi_1=3$} & $0$ & $\rm{I}_{(132)}$ &
$\rm{II}_{(12)}$ &
$\rm{III}_{(12)}$ \\[1ex]
& $1$ & $\rm{II}$        & $\rm{I}_{(13)}$  & $\rm{III}$\\[1ex]
& $2$ & $\rm{II}_{(12)}$ & $\rm{I}_{(132)}$ & $\rm{III}_{(12)}$\\[1ex]
& $3$ & $\rm{I}_{(13)}$  & $\rm{II}$        & $\rm{III}$\\[1ex]
\end{tabular}
\end{ruledtabular}
\end{table}

Unlike what happens on the quark mass matrix pair $M_u,\,M_d$, it is
remarkable to verify that the seesaw formula in Eq.~\eqref{eq:seesaw}
together with the allowed $\phi_1$, $q_3$ and $\nu_i$ charges lead
necessarily to a non-parallel structure in the leptonic mass matrix
pair $M_{e},\,m_{\nu}\,$. No NNI form was found for the effective
neutrino mass matrix $m_{\nu}$. Nevertheless, in the case where
neutrinos are Dirac fermions, a parallel structure with NNI form in
the leptonic sector is indeed possible, if one also requires the
flavour symmetry to forbid the appearance of the Majorana right-handed
mass matrix, $M_R$. The minimal discrete group realisation is
$\mathsf{Z}_7$ with, for example, the following charge assignments:
$(\phi_1,\phi_2)=(1,0)$, $\mathcal{Q}(\mathsf{10})=(2,4,3)$,
$\mathcal{Q}(\mathsf{5}^{\ast})=(3,5,4)$ and $\nu=(1,3,2)$. In this
example, the NNI mass matrix $m_D$ encodes all neutrino masses and
mixings and it was shown in Ref.~\cite{Fritzsch:2011cu} that the mass
matrix pair $M_{e},\,m_D$ with parallel structure can accommodate the
leptonic experimental data.

Analysing carefully the zero textures obtained in
Table~\ref{tab:results}, some comments are in order. If at least two
right-handed neutrinos have the same $\mathsf{Z}_4$ charge one
realises that one generation decouples from the others (i.e.,
Texture-III and its permutation $(12)\,$), it corresponds to the
right-handed neutrino charges having only values $\nu_i=0$ or $2$. Due
to the fact that the charged lepton mass matrix is in the NNI form, it
is rather easy to see that having a generation which decouples in the
neutrino sector is not phenomenologically viable, since it is
impossible to obtain large mixing angles. In a similar way Class-I,
where the effective neutrino mass matrix is a permutation of the NNI
matrix, is not viable too since it leads to small mixing angles. Also
by counting the number of independent free parameters of the mass
matrix set $M_{e},\,m_{\nu}$ in Class-I one gets a total of ten
parameters which have to account for the twelve low energy lepton
observables (six lepton masses, three mixing angles and three
phases). One is then left with only two zero textures of Class-II,
namely Texture-II and -II$_{(12)}\,$. The confrontation of such
textures with the neutrino data is explored in the next section.

\section{Numerical analysis of neutrino mass matrix within the
  Class-II}
\label{sec:neutrino}

In this section we analyse the phenomenological consequences of each
effective neutrino mass matrix belonging to the Class-II. As discussed
in the previous section only two zero textures II and
II$_{(12)}$ from Table~\ref{tab:results}, need to be confronted with
the observable neutrino data. Without loss of generality one can write
the charged lepton mass matrix, $M_{e}$, and the effective neutrino
mass matrices, $m^{(g)}_{\nu}$ of Class-\rm{II} as:
\begin{subequations}
\label{eq:lep}
\begin{align}
\label{eq:lepell}
&M_{e}=K_{e}^{\dagger}\begin{pmatrix}
    \ 0\ & \ \bar{A}_{e}(1-\epsilon_a^e)\ & \ 0\ \\
    \bar{A}_{e}(1+\epsilon_a^e) & 0 &\bar{B}_{e}(1-\epsilon_b^e)\\
    0 & \bar{B}_{e}(1+\epsilon_b^e) & C_{e}
   \end{pmatrix}\,,\\[3mm]
\label{eq:lepNu}
&m_{\nu}^{(g)}=P_g\begin{pmatrix}
\,\, 0 \,\, & \,\, A_{\nu}\,\, & 0\\
A_{\nu} & B_{\nu} & C_{\nu} \\
 0 & C_{\nu} & D_{\nu}\,e^{i\varphi}
 \end{pmatrix}P_g^{\top}\,,
\end{align}
\end{subequations}
where $g=e$ or $(12)$ according to Table~\ref{tab:results}, the
constants $\bar{A}_{e},\,\bar{B}_{e},A_{\nu},\,B_{\nu},\,C_{e,
  \nu},\,D_{\nu}$ are taken real and positive. The diagonal phase
matrix $K_{e}$ can be parameterised as
\begin{equation}
K_{e}=\diag(e^{i\kappa_1},e^{i\kappa_2},1)\,.
\end{equation}
and the phase $\varphi$ in Eq.~\eqref{eq:lepNu} cannot be absorbed by
any field redefinition. In the case of Texture-II$_{(12)}$ the
permutation matrix $P_{(12)}$ is simply given by
\begin{equation}
\label{eq:P12}
P_{(12)}=\begin{pmatrix}
\;0\;&\;1\;&\;0\;\\
1&0&0\\
0&0&1
\end{pmatrix}\,.
\end{equation}
Although the number of the parameters encoded in the pair
$M_e,m_{\nu}$ is twelve as the number of independent physical
parameters experimentally observed at low energy, the zero pattern
exhibited in Eqs.~\eqref{eq:lep} does imply new constraints among the
independent physical parameters, as it will be shown.

In order to extract the mixings angles and the CP phases encoded in
the charged lepton and the light neutrino mass matrices in
Eqs.~\eqref{eq:lep}, one needs to evaluate the
Pontecorvo-Maki-Nakagawa-Sakata (PMNS)
matrix~\cite{Pontecorvo:1957cp,Pontecorvo:1957qd,Maki:1962mu}, $U$.
It is then useful to introduce the Hermitian mass matrix $H_{e}$
defined by:
\begin{equation}
 H_{e}=M_{e}\,M^{\dagger}_{e}\,.
\end{equation}
From Eq.~\eqref{eq:lepell} one deduces that $H_{e}$ is a real Hermitian mass
matrix, which can be diagonalised by a real and orthogonal matrix, $O_{e}$,
in the following way:
\begin{equation}
 O_{e}^{\top}\,H_{e}\,O_{e}=\diag(m^2_e,\,m^2_{\mu},\,m^2_{\tau})\,.
\end{equation}
To diagonalise the effective neutrino mass matrix $m^{(g)}_{\nu}$ it is
practical to define the complex matrix $m^0_{\nu}$ as
\begin{equation}
\label{eq:mnu0}
 m^0_{\nu}=P_g^{\top}m^{(g)}_{\nu}P_g\,,
\end{equation}
where $m^0_{\nu}$ is diagonalised by $U_{\nu}$ as,
\begin{equation}
\label{eq:mnu0diag}
 U_{\nu}^{\top}\,m^0_{\nu}\,U_{\nu}=\diag(m_1,\,m_2,\,m_3)\,,
\end{equation}
and $m_i$ are the positive light neutrino masses. Finally, the PMNS
matrix $U$ is given by
\begin{equation}
\label{eq:PMNSpermut}
U=O^{\top}_{e}\,K^{\dagger}_{e}\,P_g\,U_{\nu}\,.
\end{equation}
It is useful to re-express the unitary matrix $U$ in terms of the
standard parameterisation, which
has the property to better express the neutrino observed data in a
more clear and uniform way,
\begin{widetext}
\begin{equation}
\label{eq:PMNS}
U=\begin{pmatrix}
c_{12}c_{13} & s_{12}c_{13} & s_{13}e^{-i\delta} \\
-s_{12}c_{23}-c_{12}s_{23}s_{13}e^{i\delta} &
c_{12}c_{23}-s_{12}s_{23}s_{13}e^{i\delta} &  s_{23}c_{13}\\
s_{12}s_{23}-c_{12}c_{23}s_{13}e^{i\delta} &
-c_{12}s_{23}-s_{12}c_{23}s_{13}e^{i\delta} & c_{23}c_{13}
\end{pmatrix}
\begin{pmatrix}
e^{i\alpha_1/2} & 0 & 0\\
0 & e^{i\alpha_2/2} & 0\\
0 & 0 & 1
\end{pmatrix}\,,
\end{equation}
\end{widetext}
where $s_{ij}\equiv\sin\theta_{ij}$, $c_{ij}\equiv\cos\theta_{ij}$,
$\delta$ is a Dirac CP violation phase and $\alpha_1$, $\alpha_2$ are
Majorana CP violation phases. From the neutrino oscillation
experiments one infers~\cite{Schwetz:2008er} the neutrino mass squared
differences $\Delta m^2_{21}\,,\Delta m^2_{31}$ ($\Delta
m^2_{ij}\equiv m _i^2-m _j^2$) as well as the mixing angles
$\theta_{12}\,, \theta_{23}$ and $\theta_{13}$ that are shown in
Table~\ref{tab:data}.

\begin{table}[floatfix]
\caption{\label{tab:data} The three-flavour oscillation parameters within
$1\sigma$ error from Ref.~\cite{Schwetz:2008er}.}
\begin{ruledtabular}
\begin{tabular}{cc}
parameters & $1\sigma$\\
\hline\\[-1ex]
$\Delta m^2_{21}$ &
$\left(7.59^{+0.23}_{-0.18}\right)\times10^{-5}\,\text{eV}^2$\\[2ex]
$\left|\Delta m^2_{31}\right|$
&
$\left(2.40^{+0.12}_{-0.11}\right)\times10^{-3}\,\text{eV}^2$\\[2ex]
$\sin^2\theta_{12}$
&
$0.318^{+0.019}_{-0.016}$\\[2ex]
$\sin^2\theta_{23}$
&
$0.50^{+0.07}_{-0.06}$\\[2ex]
$\sin^2\theta_{13}$
&
$<0.035$ at 90\% C.L.\\[1ex]
\end{tabular}
\end{ruledtabular}
\end{table}

In addition to oscillation parameters, we have also considered in our
numerical program three other constraints: the effective Majorana
mass, $m_{ee}$, that is proportional to the neutrinoless double beta
decay amplitude~\cite{Pascoli:2001by,Pascoli:2002xq,Pascoli:2003ke},
\begin{equation}
\label{eq:mee}
m_{ee}\equiv\sum_{i=1}^3 m_i\,U^{\ast2}_{1i}\,,
\end{equation}
the constraint from Tritium $\beta$ decay~\cite{Nakamura:2010zzi}, $m_{\nu_e}$,
\begin{equation}
\label{eq:tritium}
m_{\nu_e}^2\equiv\sum_{i=1}^3\,m_i^2|U_{1i}|^2<\left(2.3\,\text{eV}
\right)^2\quad\text{at 95\% C.L.}\,,
\end{equation}
and the bound on the sum of light neutrino masses, $\mathcal{T}$ from
cosmological and astrophysical data:
\begin{equation}
\label{eq:cosmo}
\mathcal{T}\equiv\sum_{i=1}^3 m_i < 0.68\,\text{eV}\quad\text{at 95\% C.L.}\,.
\end{equation}
This upper limit on $\mathcal{T}$ results from the combination of the
Cosmic Microwave Background data of the WMAP experiment with
supernovae data and data on galaxy
clustering~\cite{Spergel:2006hy}. 

Bounds on $|m_{ee}|$ can be estimated by taking into account the best
fit values of neutrino oscillation parameters and the upper limit of
$\sin^2\theta_{13}$ from Table~\ref{tab:data} together with the
assumption of a hierarchical spectrum for the light neutrino
masses~\cite{Bilenky:2001rz,Bilenky:2001xq,Petcov:2005yq},
\begin{subequations}
\label{eq:meebound}
\begin{equation}
\label{eq:meeNH}
|m_{ee}|\lesssim0.005\,\text{eV}\,\text{(NH)}\,,
\end{equation}
\begin{equation}
\label{eq:meeIH}
10^{-2}\,\text{eV}\lesssim|m_{ee}|\lesssim0.05\,\text{eV}\,\text{(IH)}\,.
\end{equation}
\end{subequations}

Since neutrino oscillations are only sensitive to the mass squared
differences, the constraints given by Eqs.~\eqref{eq:tritium} and
\eqref{eq:cosmo} can determine the lightest neutrino mass: $m_1$ in
the case of normal hierarchy (NH) as in the charged lepton masses or
$m_3$ in the case of inverted hierarchy (IH). The fact that these two
hierarchies are still phenomenologically viable is due to the
indetermination on the sign of the mass squared difference $\Delta
m^2_{31}$. The constraint given by Eq.~\eqref{eq:tritium} implies that
$m_1$ (NH) or $ m_3$ (IH) should be less than $2.3$~eV, which is a
rather poor constraint and needs to get an improvement in future
experiments.

The cosmological bound is much more severe and even if one takes the
upper bound of quantity $\mathcal{T}$ given in Eq.~\eqref{eq:cosmo},
one gets
\begin{equation}
\label{eq:cosmoslight}
m_1<0.22\,\text{eV (NH)}\,,\qquad
m_3<0.22\,\text{eV (IH)}\,.
\end{equation}
If one refers to a more restrictive bound on $\mathcal{T}\leq0.28$ eV
at 95\% C.L.~\cite{Thomas:2009ae}, one then gets for the lightest
neutrino mass the following restrictions:
\begin{equation}
\label{eq:cosmoslight2}
m_1<0.089\,\text{eV (NH)}\,,\qquad
m_3<0.084\,\text{eV (IH)}\,.
\end{equation}

The real and orthogonal matrix $O_{e}$ can be expressed in terms of
the parameters $\epsilon_{a,b}^{e}$ defined in Eq.~\eqref{eq:eps} and
the charged lepton masses are
\begin{subequations}
\label{eq:lepmasses}
\begin{align}
  m_e&=0.486661305\pm{0.000000056}\text{ MeV}\,,\\
  m_{\mu}&=102.728989\pm{0.000013}\text{ MeV}\,,\\
  m_{\tau}&=1746.28\pm{0.16}\text{ MeV}\,,
 \end{align}
\end{subequations}
evaluated at $M_Z$ scale through the renormalisation group equations
for QED in the $\overline{MS}$ scheme at 1-loop
level~\cite{Fusaoka:1998vc,Xing:2007fb}.  In the limit where the
parameters $\epsilon_{a,b}^{e}$ are small, the orthogonal matrix
$O_{e}$ that diagonalises $M_{e}$ takes
approximately~\cite{Branco:1992ba,Branco:2010tx} the following form:
\begin{subequations}
\label{eq:Ol}
\begin{align}
(O_{e})_{12}&\approx-\sqrt{\frac{m_e}{m_{\mu}}}\left(1-\epsilon^{e}_a-\frac{m_{\mu}}{m_{\tau}}\epsilon^{e}_b\right)\,,\\
(O_{e})_{13}&\approx\sqrt{\frac{m_e\,m_{\mu}^2}{m_{\tau}^3}}\left(1+\epsilon^{e}_b-\epsilon^{e}_a\right)\,,\\
(O_{e})_{21}&\approx\sqrt{\frac{m_e}{m_{\mu}}}\left(1-\epsilon^{e}_a-\frac{m_e}{m_{\tau}}\epsilon^{e}_b\right)\,,\\
(O_{e})_{23}&\approx \sqrt{\frac{m_{\mu}}{m_{\tau}}}\left(1-\epsilon^{e}_b\right)\,,\\
(O_{e})_{31}&\approx-\sqrt{\frac{m_e}{m_{\tau}}}\left(1-\epsilon^{e}_a-\epsilon^{e}_b\right)\,,\\
(O_{e})_{32}&\approx-\sqrt{\frac{m_{\mu}}{m_{\tau}}}\left(1-\epsilon^{e}_b+\frac{m_e}{m_{\mu}}\epsilon^{e}_a\right)\,.
\end{align}
\end{subequations}

The unitary matrix, $U_{\nu}$, that diagonalises the neutrino mass
matrix $m^0_{\nu}$ given in Eq.~\eqref{eq:mnu0} can be written in
terms of the neutrino masses, the real parameter $D_{\nu}$ and the
phase $\varphi$ from Eq.~\eqref{eq:lepNu}. If one takes the phase
$\varphi=0$ the matrix $U_{\nu}$ becomes an exact form of the
remaining parameters~\cite{Branco:2007nn}, with the matrix element
moduli,
\begin{subequations}
\label{eq:13sym}
\begin{align}
|(U_{\nu})_{11}|&=\sqrt{\frac{m_2 m_3 (D_{\nu}-m_1)}{D_{\nu}\,
    (m_2-m_1) (m_3-m_1)}}\,,\\
|(U_{\nu})_{12}|&=\sqrt{\frac{m_1 m_3 (m_2-D_{\nu})}{D_{\nu}\,
    (m_2-m_1) (m_3-m_2)}}\,,\\
|(U_{\nu})_{13}|&=\sqrt{\frac{m_1 m_2 (D_{\nu}-m_3)}{D_{\nu}\,
    (m_3-m_1) (m_3-m_2)}}\,,\\
|(U_{\nu})_{21}|&=\sqrt{\frac{m_1 (m_1-D_{\nu})}{(m_2-m_1)
    (m_3-m_1)}}\,,\\
|(U_{\nu})_{22}|&=\sqrt{\frac{(D_{\nu}-m_2) m_2}{(m_2-m_1)
    (m_3-m_2)}}\,,\\
|(U_{\nu})_{23}|&=\sqrt{\frac{m_3 (m_3-D_{\nu})}{(m_3-m_1)
    (m_3-m_2)}}\,,\\
|(U_{\nu})_{31}|&=\sqrt{\frac{m_1 (D_{\nu}-m_2)
    (D_{\nu}-m_3)}{D_{\nu}\, (m_2-m_1) (m_3-m_1)}}\,, \\
|(U_{\nu})_{32}|&=\sqrt{\frac{m_2 (D_{\nu}-m_1)
    (m_3-D_{\nu})}{D_{\nu}\, (m_2-m_1) (m_3-m_2)}}\,,\\
|(U_{\nu})_{33}|&=\sqrt{\frac{m_3 (D_{\nu}-m_1)
    (D_{\nu}-m_2)}{D_{\nu}\, (m_3-m_1) (m_3-m_2)}}\,,
\end{align}
\end{subequations}
and the signs for the matrix elements $(U_{\nu})_{ij}$ are in the case
of normal hierarchy given by
\begin{equation}
\begin{pmatrix}
+ & + & +\\
\sign(m_1) & \sign(m_2) & \sign(m_3)\\
\sign(m_2 m_3) & \sign(m_1) & +
\end{pmatrix}\,,
\end{equation}
and in the case of inverted hierarchy by
\begin{equation}
\begin{pmatrix}
+ & + & +\\
\sign(m_1) & \sign(m_2) & \sign(m_3)\\
\sign(m_3) & + & \sign(m_1 m_2)
\end{pmatrix}\,.
\end{equation}
In our numerics we have performed the full diagonalisation with all
possible values for $\varphi$ by just applying the
Eq.~\eqref{eq:mnu0diag}.

To study the new constraints that arise from a charged lepton mass
matrix $M_e$ in the NNI form and the effective neutrino mass matrix
$m_{\nu}$ belonging to Class-II, we have varied all experimental
charged lepton masses and neutrino mass differences within their
allowed range given in Eq.~\eqref{eq:lepmasses} and
Table~\ref{tab:data}, respectively.  The mass of the lightest neutrino
($m_1$ in NH or $m_3$ in IH) was scanned for different magnitudes
below 2 eV. In order to fully reconstruct the PMNS matrix, we have
also varied the free parameters $\epsilon^{e}_{a,b}$, $D_{\nu}$ and
the phases $\kappa_1,\,\kappa_2,\,\varphi$, defined in
Eq.~\eqref{eq:lep}. The remaining six parameters,
$\bar{A}_{e},\,\bar{B}_{e},A_{\nu},\,B_{\nu},\,C_{e, \nu}$ are
determined from the values of lepton masses and $\epsilon^{e}_{a,b}$,
$D_{\nu}$, $\varphi$.  The restriction in this scan was to accept only
the input values which correspond to a reconstructed PMNS matrix~$U$
that naturally leads to the mixing
angles $\theta_{12}\,,\theta_{23}$ and $\theta_{13}$ within their
experimental bounds presented in Table~\ref{tab:data}.

\begin{figure}[ht]
\begin{center}
\includegraphics[width=8.6cm,clip]{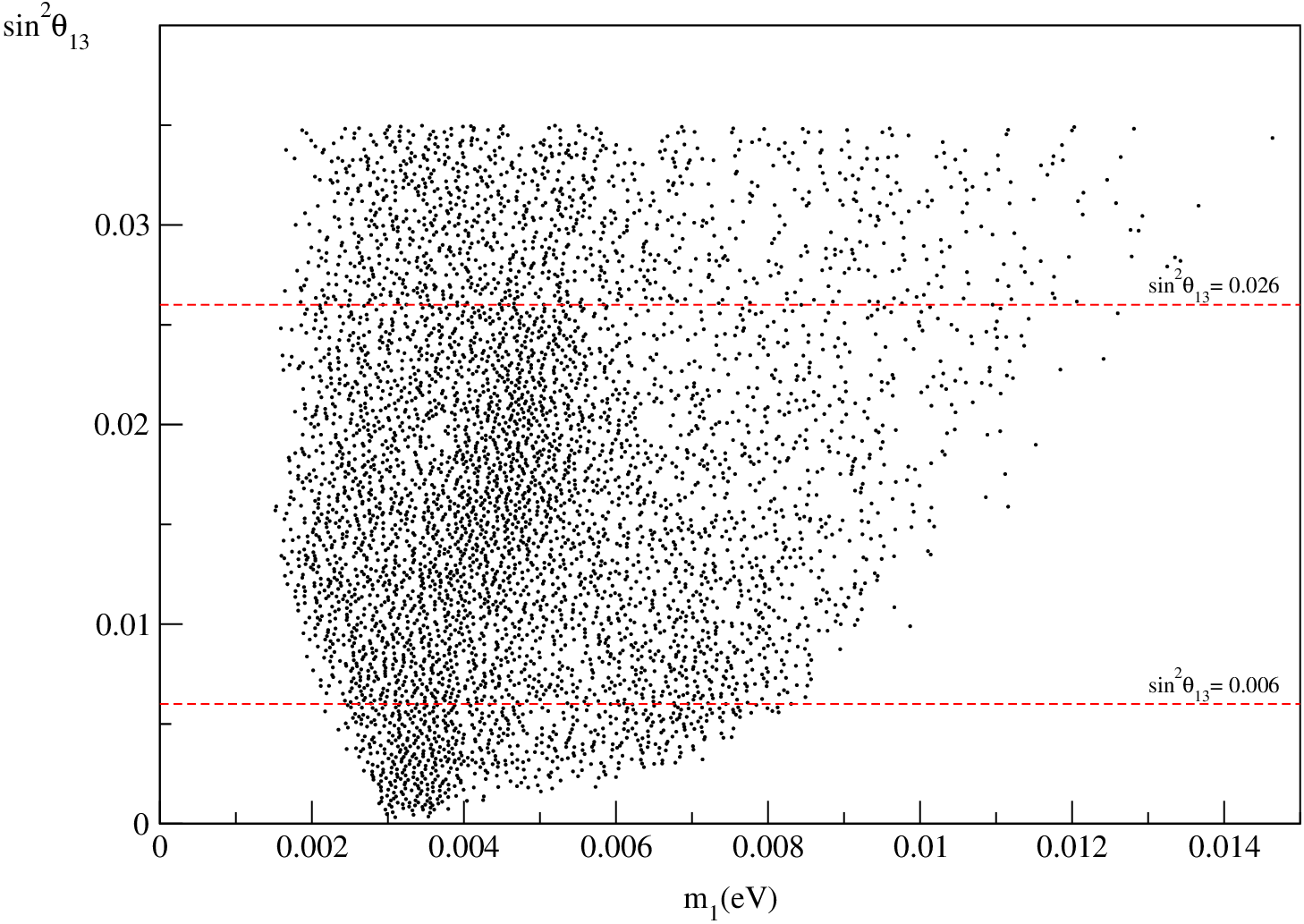}
\end{center}
\caption{\label{fig2} Plot of $\sin^2 \theta_{13}$ as a function of
  $m_1$ in the case of Texture-II and normal hierarchy. The dashed
  lines correspond to $\sin^2\theta_{13}=0.016\pm0.010$ at 1$\sigma$
  C.L. from the global analysis~\cite{Fogli:2008jx}. The plot of
  $\sin^2 \theta_{23}$ vs $m_1$ is not present here, since it reveals
  no correlation.}
\end{figure}

\begin{figure}[ht]
\begin{center}
\includegraphics[width=9cm,clip]{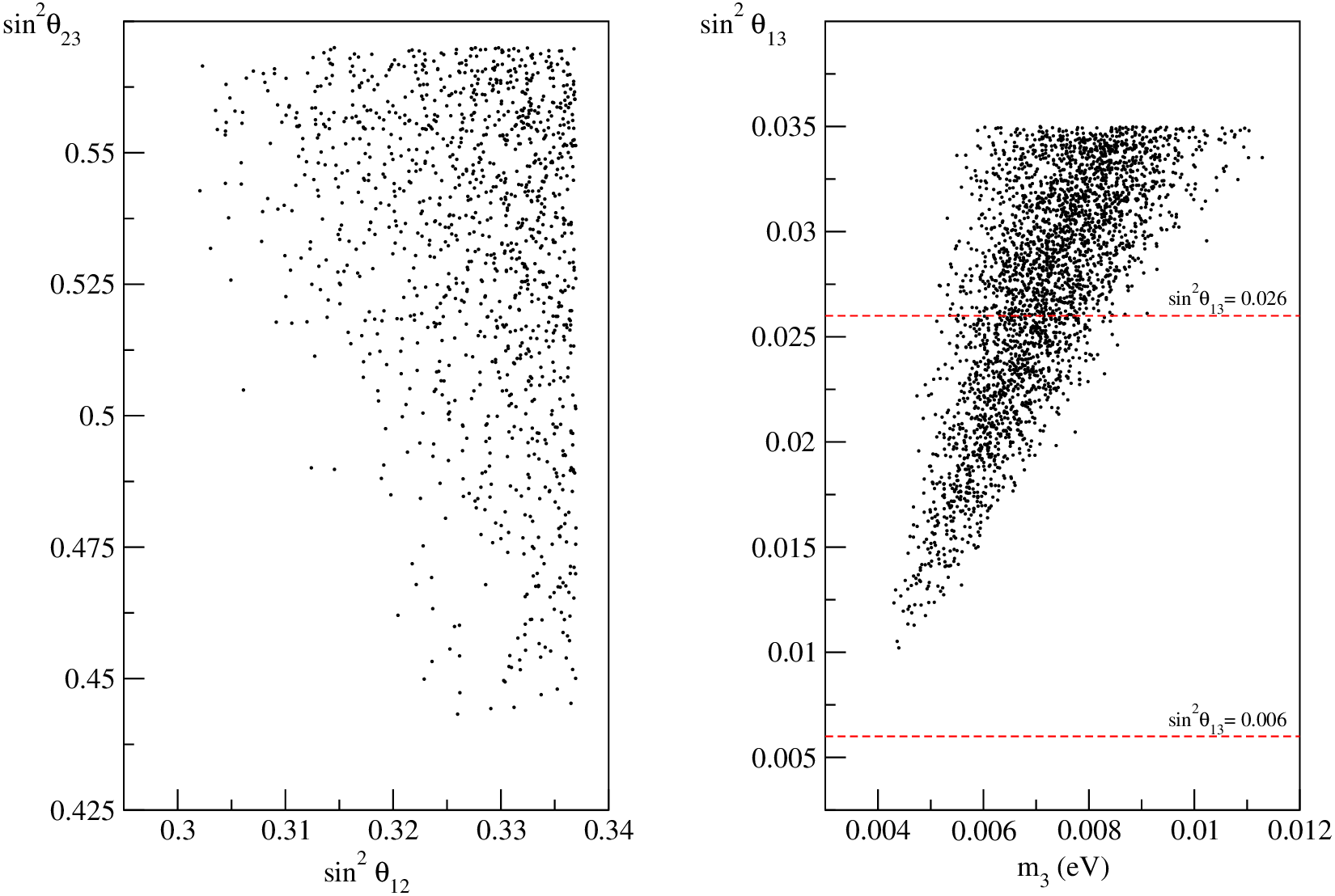}
\end{center}
\caption{\label{fig3} Plots of $\sin^2 \theta_{23}$ versus
  $\sin^2\theta_{12}$~(left) and $\sin^2 \theta_{13}$ versus
  $m_3$~(right) in the case of Texture-II$_{(12)}$ and inverted
  hierarchy. The dashed lines correspond to
  $\sin^2\theta_{13}=0.016\pm0.010$ at 1$\sigma$ C.L. from the global
  analysis~\cite{Fogli:2008jx}.}
\end{figure}

\begin{figure}[ht]
\begin{center}
\includegraphics[width=9cm,clip]{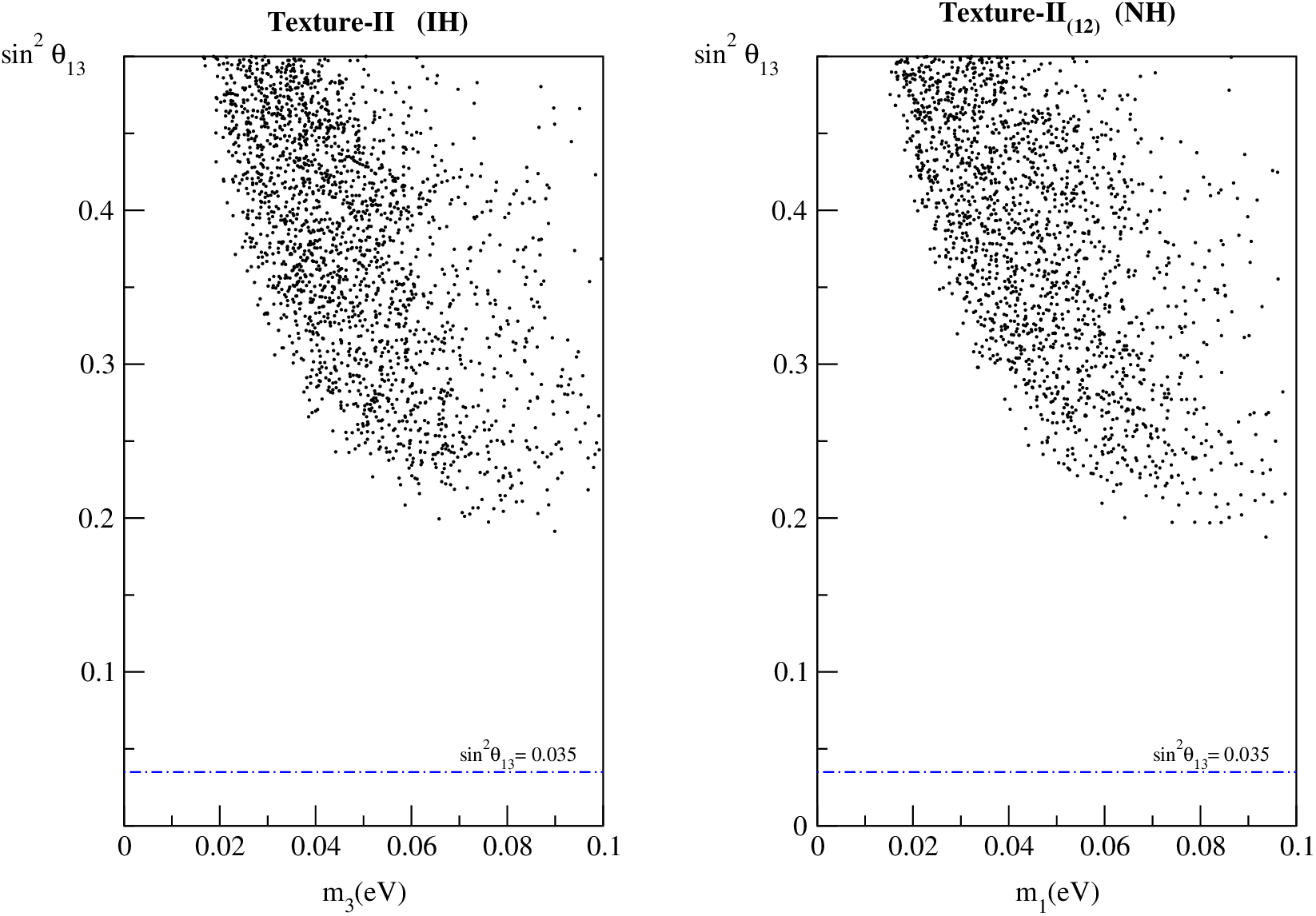}
\end{center}
\caption{\label{fig4} Plots of $\sin^2 \theta_{13}$ as a function of
  the lightest neutrino mass for Texture-II (left), $m_3$ for inverted
  hierarchy, and Texture-II$_{(12)}$ (right), $m_1$ for normal
  hierarchy. The dashed-pointed line corresponds to the limit
  $\sin^2\theta_{13}<0.035$ at 90\% C.L. from the global
  analysis~\cite{Schwetz:2008er}.}
\end{figure}

From our search, we have found that the neutrino mass hierarchy can
distinguish the two zero textures, II and II$_{(12)}$, since
Texture-II is only compatible with normal hierarchy, while
Texture-II$_{(12)}$ requires inverted hierarchy. In Fig.~\ref{fig2},
we plot $\sin^2\theta_{13}$ against the lightest neutrino mass $m_1$
for the Texture-II where normal hierarchy applies. For the mixing
angles $\theta_{12}$ and $\theta_{23}$ we found no correlation at
all. While in Fig.~\ref{fig3}, we plot $\sin^2\theta_{23}$,
$\sin^2\theta_{13}$ over the lightest neutrino mass, which is $m_3$
for the Texture-II$_{(12)}$ since it is only compatible with inverted
hierarchy. In the plots of $\sin^2\theta_{13}$ over the lightest
neutrino mass shown in Fig.~\ref{fig2} and~\ref{fig3} we have also
drawn the new hint of non-zero $\sin^2\theta_{13}$:
\begin{equation}
\label{eq:t13bound}
 \sin^2\theta_{13}=0.016\pm0.010\,,
\end{equation}
at 1$\sigma$ C.L. from the global analysis~\cite{Fogli:2008jx} of all
available neutrino oscillation data. A direct determination of
$\sin^2\theta_{13}$ can affect significantly the validity of our
model. It is indeed clear in Fig.~\ref{fig4} that neither Texture-II
nor Texture-II$_{(12)}$ are compatible with inverted or normal
hierarchy, respectively. In both situations, the value of $|U_{13}|$
is two orders of magnitude outside the required bound for
$\sin^2\theta_{13}\,$ given in Table~\ref{tab:data}.

\begin{figure}[ht]
\begin{center}
\includegraphics[width=9cm,clip]{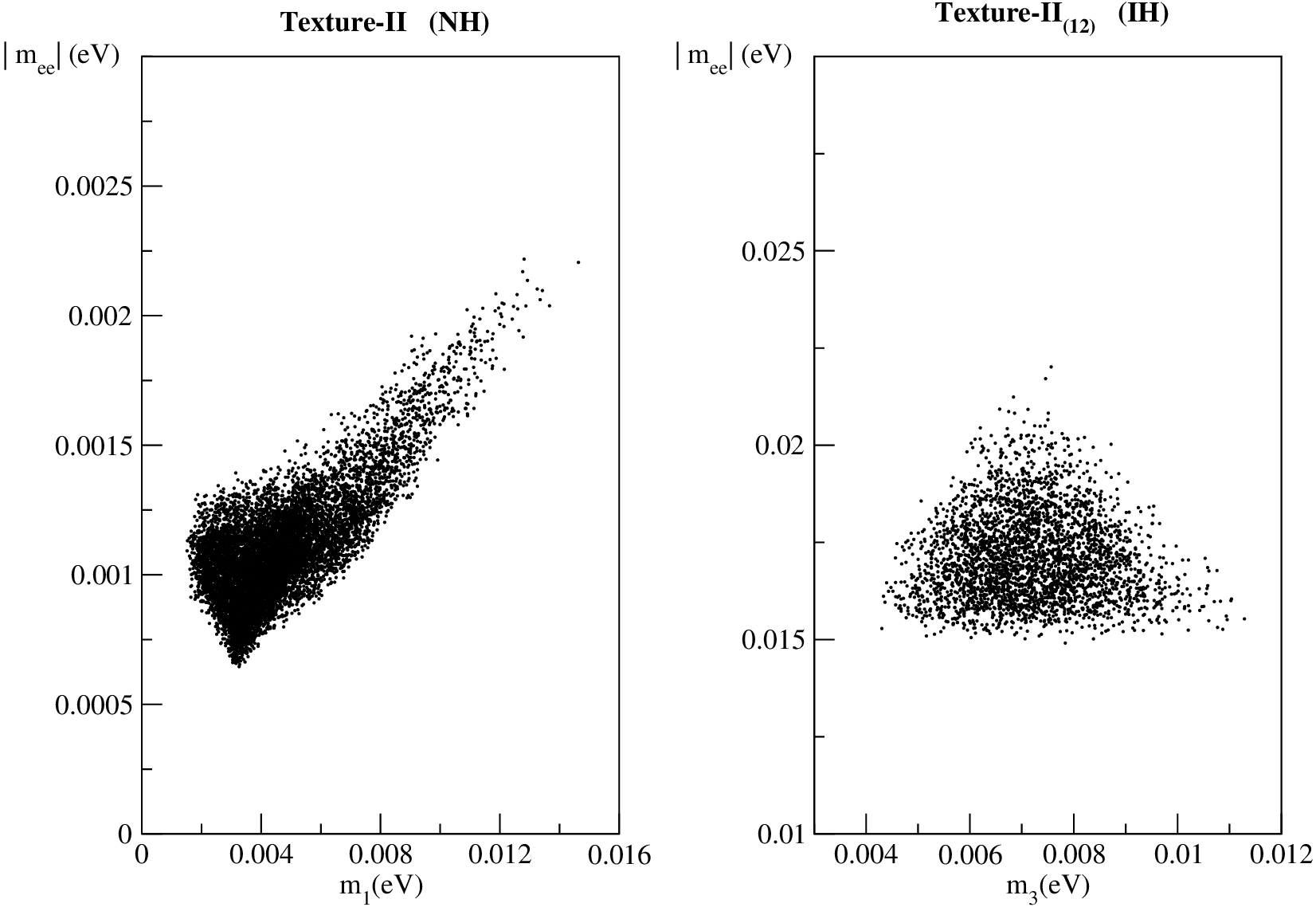}
\end{center}
\caption{\label{fig5} Plotting $|m_{ee}|$ versus
$m_1$~(left) and $m_3$~(right)
for Texture-II (NH) and -II$_{(12)}$ (IH), respectively.}
\end{figure}

One can clearly see from Fig.~\ref{fig2} and Fig.~\ref{fig3} that for
both textures the lightest neutrino mass, $m_1$ (NH) or $m_3$ (IH), is
bounded. In the case of Texture-II one has,
\begin{equation}
\label{eq:m1lim}
 0.0013\,\text{eV}\leq m_1\leq 0.016\,\text{eV}\,,
\end{equation}
whereas for Texture-II$_{(12)}$ one has,
\begin{equation}
\label{eq:m3lim}
0.0042\,\text{eV}\leq m_3\leq 0.011\,\text{eV}\,,
\end{equation}
which let us to conclude that the light neutrino mass spectrum cannot
be quasi-degenerated and there is no room to account for a massless
neutrino state (as it is still allowed by general neutrino oscillation
analyses). The upper bound on the constraint from Tritium $\beta$
decay given in Eq.~\eqref{eq:tritium} does not have any impact in these
results. Although the cosmological and astrophysical upper bound on
the sum of light neutrino masses seems rather severe, the upper limit
on the lightest neutrino mass given in Eq.~\eqref{eq:cosmoslight} or
even the most restrictive value in Eq.~\eqref{eq:cosmoslight2} are
indeed above the bounds reported in Eqs.~\eqref{eq:m1lim}
and~\eqref{eq:m3lim}.

Concerning the effective Majorana mass $|m_{ee}|$, we present in
Fig.~\ref{fig5} the value of $|m_{ee}|$ as function of the lightest
neutrino mass for both Texture-II and -II$_{(12)}$. From our scan, we
have obtained the following limits for $|m_{ee}|\,$:
\begin{equation}
6.4\times10^{-4}\,\text{eV}<|m_{ee}|<2.2\times10^{-3}\,\text{eV}\,,
\end{equation}
in the case of Texture-II and,
\begin{equation}
 0.015\,\text{eV}<|m_{ee}|<0.022\,\text{eV}\,,
\end{equation}
for Texture-II$_{(12)}\,$. These bounds obtained for $|m_{ee}|$ are in
full agreement with those given in Eq.~\eqref{eq:meebound}. One expect
that future improvements on the experimental value of $|m_{ee}|$ may
have impact on this textures (II and II$_{(12)}$) or on the
$\mathsf{Z}_4$ symmetry itself, since a change on the magnitude of the
experimental value of $|m_{ee}|$ may drastically constrains our model.

Before closing the section one may address the question which could be
the smallest deviations to the Fritzsch ansatz, $\varepsilon_e$, for
the charged lepton mass matrix $M_e$ acceptable by the experimental
data as it was done for the quark sector in
Ref.~\cite{Branco:2010tx}. In the quark sector, the lower bound for
$\varepsilon_q$ consistent with electroweak data can be evaluated by
taking into account the values of the quark masses, CKM elements
($V_{us}\,,V_{cb}\,,V_{ub}$) and the angle
$\beta\equiv\arg(-V_{cd}V_{cb}^{\ast}V^{\ast}_{td}V_{tb})\,$ of the
unitarity triangle listed in Table~\ref{tab:SM}. We have performed an
update of the lower bound of the parameter $\varepsilon_q$ defined in
Eq.~\eqref{eq:eps}, \begin{equation} \label{eq:epsqbound}
  \varepsilon_q\geq0.188\,, \end{equation} which has increased more
than 10\% compared with the value reported in~\cite{Branco:2010tx}.

\begin{table}[ht]
  \caption{\label{tab:SM} Values of the quark masses,
    the CKM element
    moduli $|V_{us}|\,$, $|V_{cb}|\,$, $|V_{ub}|$ and the angle
    $\beta$ of the unitarity
    triangle at the scale
    $M_Z$. The quark masses are
    calculated at $M_Z$ scale by taking properly into account the 4-loop
    renormalisation group equations for QCD in the $\overline{MS}$
    scheme~\cite{Rodrigo:1993hc,Chetyrkin:1996ia,vanRitbergen:1997va,
      Fusaoka:1998vc,Xing:2007fb} and using as input the masses given
    in~\cite{Nakamura:2010zzi}.}
\begin{ruledtabular}
\begin{tabular}{c}
$\begin{array}{l@{\hspace*{12mm}}l}
m_u=1.4\pm{0.5}\,\text{MeV} &
m_d=2.9\pm{0.5}\,\text{MeV}\\[2mm]
m_c=0.62^{+0.06}_{-0.07}\,\text{GeV}  &
m_s=58^{+16}_{-12}\,\text{MeV} \\[2mm]
m_t=170.2\pm{1.0}\,\text{GeV} &
m_b=2.86^{+0.16}_{-0.06}\,\text{GeV}
\\[4mm]
\end{array}
$\\
$\begin{array}{l@{\hspace*{10mm}}l}
|V_{us}|=0.2253\pm0.0007 &
 |V_{ub}|=\left(3.47^{+0.16}_{-0.12}\right)\times10^{-3}\\[2mm]
 |V_{cb}|=\left(41.0^{+1.1}_{-0.7}\right)\times10^{-3} &
 \sin2\beta=0.673\pm0.023
\end{array}
$
\end{tabular}
\end{ruledtabular}
\end{table}

The deviation from Hermiticity of the charged lepton mass matrix
$\varepsilon_{e}$ was computed for the Texture-II and -II$_{(12)}$ and
the lower bounds $\varepsilon_{e}>0.0011$ and
$\varepsilon_{e}>0.0013\,$ were found, respectively. These bounds
imply that the neutrino data allow for the charged lepton mass
matrices $M_{e}$ to be much closer to Hermiticity than the quark mass
matrices, by two orders of magnitude higher.


\section{Conclusions}
\label{sec:conclusions}

In this work we extended the flavour symmetry proposed in
Ref.~\cite{Branco:2010tx} in the context of $\mathsf{SU(5)}$ Grand
Unification.  Such symmetry was imposed in the Lagrangian in order to
force the quark mass matrices $M_u,\,M_d$ to have NNI form, after
spontaneous gauge symmetry breaking. In this GUT model, beyond the
standard field content, three right-handed neutrinos and two Higgs
quintets were added. It turns out that the light neutrino masses are
generated through type-I seesaw mechanism and the low energy theory
below the GUT scale is just a two Higgs doublet model. In this
context, the minimal realisation of such discrete symmetry is
$\mathsf{Z}_4$.

Since in the $\mathsf{SU(5)}\times\mathsf{Z}_4$ model the charged
lepton mass matrix, $M_{e}$, are related to the down-quark mass matrix
$M_d$, it was not surprising to verify that the mass matrix $M_{e}$
acquired a NNI form, too.  Instead, the right-handed neutrinos being
singlets under $\mathsf{SU(5)}$, their $\mathsf{Z}_4$ charges are free
parameters and the neutrino mass matrix, $m_{\nu}$ has no NNI
form. There are in fact only six zero textures allowed for the
effective neutrino mass matrix, $m_{\nu}$. However, the neutrino
oscillation data select just two textures among the six possibilities:
Texture-II, which is compatible only with normal hierarchy for the
neutrino mass spectrum and the Texture-II$_{(12)}$, which demands
neutrino mass spectrum to have inverted hierarchy.

\begin{table}[ht]
\caption{\label{tab:conclusions} Summary of Texture-II and -II$_{(12)}$
results.}
\begin{ruledtabular}
\begin{tabular}{c@{\hspace{4mm}}c}
\textbf{Texture-II (NH)}&\textbf{Texture-II$_{(12)}$ (IH)}\\[3mm]
 $\begin{pmatrix}
  \;0\; & \;\star\; & \;0\; \\
  \star & \star & \star\\
  0 & \star & \star
 \end{pmatrix}$
& $\begin{pmatrix}
  \;\star\; & \;\star\; & \;\star\;\\
  \star & 0 & 0\\
  \star & 0 & \star
 \end{pmatrix}$\\ \\
&$\sin^2\theta_{13}>0.010$\\[2mm]
$0.0013\,\text{eV} \leq m_1  \leq 0.016\,\text{eV}$
& $0.0042\,\text{eV}\leq m_3\leq 0.011\,\text{eV}$\\[2mm]
$0.00064\,\text{eV} <|m_{ee}| <0.0022\,\text{eV}$
&$0.015\,\text{eV}<|m_{ee}|<0.022\,\text{eV}$\\[2mm]
$\varepsilon_{e}>0.0011$
& $\varepsilon_{e}>0.0013$
\\[1mm]
\end{tabular}
\end{ruledtabular}
\end{table}

In Table~\ref{tab:conclusions}, we summarise the main predictions for
Texture-II and -II$_{(12)}$. In both textures, the lightest neutrino
mass is predicted to be bounded and the neutrino mass spectrum to be
hierarchical, therefore not compatible with a massless neutrino. Our results
are also in agreement with three other relevant constraints, namely
the effective Majorana mass $m_{ee}$, the constraint from Tritium $\beta$ decay
and the cosmological bound on the sum of light neutrino masses. We
have also obtained that the deviation of Hermiticity on the charged
lepton mass matrices is two orders of magnitude lower than the one
estimated for the quark sector.

Future improvements on the knowledge of neutrino oscillations, neutrinoless double beta
decay, tritium beta decay and cosmological astrophysics measurements
may be decisive for testing the viability of the
$\mathsf{SU(5)}\times\mathsf{Z}_4$ model.


\begin{acknowledgments}
  We would like to thank Gustavo C. Branco and M. N. Rebelo for
  fruitful discussions and reading carefully the manuscript. We would
  also like to thank Sergio Palomares-Ruiz for pointing out some
  useful remarks. This work was partially supported by Funda\c{c}\~ao
  para a Ci\^encia e a Tecnologia (FCT, Portugal) through the projects
  CERN/FP/109305/2009, CERN/FP/116328/2010, PTDC/FIS/098188/2008 and
  CFTP-FCT Unit 777 which are partially funded through POCTI (FEDER),
  by Marie Curie Initial Training Network ”UNILHC”
  PITN-GA-2009-237920, by Accion Complementaria Luso-Espanhola FCT and
  MICINN with project number 20NSML3700. The work of C.~Sim\~oes is
  also supported by FCT under the contract SFRH/BD/61623/2009.
\end{acknowledgments}


\appendix

\section{Two-loop evolution of the running gauge couplings}
\label{a:beta}

In this appendix we collect the two-loop renormalisation group
equations for the gauge coupling constants $\alpha_i \, (i=1,2,3)$,
which can be written in the
form~\cite{Jones:1981we,Machacek:1983tz,Luo:2002ti}
\begin{equation}
\begin{aligned}
\label{eq:rge}
\frac{d}{dt}\alpha^{-1}_i&=-\frac{b_i}{2\pi}
- \frac{1}{8\pi^2}\sum_{j} b_{ij}
\alpha_j\\
&+\frac{1}{32\pi^3}\sum_{\substack{f=u,d,e,\\k=1,2}}C_{if}\,
\text{Tr}
\left(\Gamma_f^{k\,\dagger}\Gamma_f^k\right)\,,
\end{aligned}
\end{equation}
where $\alpha_1 = 5/3\,\alpha_y$, $b_i$ are the usual one-loop beta
coefficients, $b_{ij}$ and $C_{if}$ are the two-loop beta
coefficients. The quantities $\Gamma^{1,2}_f$ denote the quark and
lepton Yukawa coupling matrices corresponding to the Higgs doublets
$\Phi_1,\Phi_2$. At the unification scale $\Lambda$, the gauge
couplings $\alpha_i$ obey to the relation
\begin{equation}
\alpha_1(\Lambda) = \alpha_2(\Lambda)= \alpha_3(\Lambda)\,.
\end{equation}

In the region of energy scales where one has only the SM degrees of
freedom, the $\beta$-function coefficients are given by:
\begin{equation}
b_i=\begin{pmatrix}
\frac{41}{10} \\ -\frac{19}{6} \\ -7\end{pmatrix}\,,\,
b_{ij}=\begin{pmatrix}
\frac{199}{50} & \frac{27}{10} & \frac{44}{5}\\[1.5mm]
\frac{9}{10} & \frac{35}{6} & 12\\[1.5mm]
\frac{11}{10} & \frac{9}{2} & -26
\end{pmatrix}\,.
\end{equation}
The two-loop coefficients $C_{if}$ are given by,
\begin{equation}
C_{if}=\begin{pmatrix}
\frac{17}{10} & \frac{1}{2} & \frac{3}{2}\\[1.5mm]
\frac{3}{2} & \frac{3}{2} & \frac{1}{2}\\[1.5mm]
2 & 2 & 0
\end{pmatrix}\,,
\end{equation}
and they are neglected in our Runge-Kutta integration.

Concerning the $\beta$-function coefficients for the relevant particle
content of the $\mathsf{SU(5)}$ theory, absent in the SM, one has the
following coefficients for the doublets $\Phi_1,\Phi_2$, the colour
triplets $T_1,T_2$, the triplet $\Sigma_3$ and the octet $\Sigma_8$:
\begin{align}
b_i^{\Phi_{1,2}}&=\begin{pmatrix}
\frac{1}{10} \\ \frac{1}{6} \\ 0
\end{pmatrix}\,,
\qquad
b_{ij}^{\Phi_{1,2}}=\begin{pmatrix}
\frac{9}{50} &\frac{9}{10} & 0\\[1.5mm]
\frac{3}{10} &\frac{13}{6} & 0\\[1.5mm]
0 & 0 & 0
\end{pmatrix}\,,
\\
b_i^{T_{1,2}}&=\begin{pmatrix}
\frac{1}{15} \\ 0 \\ \frac{1}{6}
\end{pmatrix}\,,
\qquad
b_{ij}^{T_{1,2}}=\begin{pmatrix}
\frac{4}{75} & 0 & \frac{16}{15}\\[1.5mm]
0 & 0 & 0\\[1.5mm]
\frac{2}{15} & 0 & \frac{11}{3}
\end{pmatrix}\,,
\\
b_i^{\Sigma_3}&=\begin{pmatrix}
0 \\ \frac{2}{3} \\ 0
\end{pmatrix}\,,
\qquad
b_{ij}^{\Sigma_3}=\begin{pmatrix}
0 & 0 & 0\\[1.5mm]
0 & \frac{56}{3} & 0\\[1.5mm]
0 & 0 & 0
\end{pmatrix}\,,
\\
b_i^{\Sigma_8}&=\begin{pmatrix}
0 \\ 0 \\ 1
\end{pmatrix}\,,
\qquad
b_{ij}^{\Sigma_8}=\begin{pmatrix}
0 & 0 & 0\\[1.5mm]
0 & 0 & 0\\[1.5mm]
0 & 0 & 42
\end{pmatrix}\,,
\end{align}
which are introduced at the appropriate intermediate scales.


\bibliography{refs}

\begin{thebibliography}{63}%
\makeatletter
\providecommand \@ifxundefined [1]{%
 \@ifx{#1\undefined}
}%
\providecommand \@ifnum [1]{%
 \ifnum #1\expandafter \@firstoftwo
 \else \expandafter \@secondoftwo
 \fi
}%
\providecommand \@ifx [1]{%
 \ifx #1\expandafter \@firstoftwo
 \else \expandafter \@secondoftwo
 \fi
}%
\providecommand \natexlab [1]{#1}%
\providecommand \enquote  [1]{``#1''}%
\providecommand \bibnamefont  [1]{#1}%
\providecommand \bibfnamefont [1]{#1}%
\providecommand \citenamefont [1]{#1}%
\providecommand \href@noop [0]{\@secondoftwo}%
\providecommand \href [0]{\begingroup \@sanitize@url \@href}%
\providecommand \@href[1]{\@@startlink{#1}\@@href}%
\providecommand \@@href[1]{\endgroup#1\@@endlink}%
\providecommand \@sanitize@url [0]{\catcode `\\12\catcode `\$12\catcode
  `\&12\catcode `\#12\catcode `\^12\catcode `\_12\catcode `\%12\relax}%
\providecommand \@@startlink[1]{}%
\providecommand \@@endlink[0]{}%
\providecommand \url  [0]{\begingroup\@sanitize@url \@url }%
\providecommand \@url [1]{\endgroup\@href {#1}{\urlprefix }}%
\providecommand \urlprefix  [0]{URL }%
\providecommand \Eprint [0]{\href }%
\@ifxundefined \urlstyle {%
  \providecommand \doi  [0]{\begingroup \@sanitize@url \@doi}%
  \providecommand \@doi [1]{\endgroup \@@startlink {\doibase
  #1}doi:\discretionary {}{}{}#1\@@endlink }%
}{%
  \providecommand \doi  [0]{doi:\discretionary{}{}{}\begingroup
  \urlstyle{rm}\Url }%
}%
\providecommand \doibase [0]{http://dx.doi.org/}%
\providecommand \Doi [0]{\begingroup \@sanitize@url \@Doi }%
\providecommand \@Doi  [1]{\endgroup\@@startlink{\doibase#1}\@@Doi}%
\providecommand \@@Doi [1]{#1\@@endlink}%
\providecommand \selectlanguage [0]{\@gobble}%
\providecommand \bibinfo  [0]{\@secondoftwo}%
\providecommand \bibfield  [0]{\@secondoftwo}%
\providecommand \translation [1]{[#1]}%
\providecommand \BibitemOpen [0]{}%
\providecommand \bibitemStop [0]{}%
\providecommand \bibitemNoStop [0]{.\EOS\space}%
\providecommand \EOS [0]{\spacefactor3000\relax}%
\providecommand \BibitemShut  [1]{\csname bibitem#1\endcsname}%
\bibitem [{\citenamefont {Georgi}\ and\ \citenamefont
  {Glashow}(1974)}]{Georgi:1974sy}%
  \BibitemOpen
  \bibfield  {author} {\bibinfo {author} {\bibfnamefont {H.}~\bibnamefont
  {Georgi}}\ and\ \bibinfo {author} {\bibfnamefont {S.}~\bibnamefont
  {Glashow}},\ }\Doi {10.1103/PhysRevLett.32.438} {\bibfield  {journal}
  {\bibinfo  {journal} {Phys.Rev.Lett.},\ }\textbf {\bibinfo {volume} {32}},\
  \bibinfo {pages} {438} (\bibinfo {year} {1974})}\BibitemShut {NoStop}%
\bibitem [{\citenamefont {Dorsner}\ and\ \citenamefont
  {Fileviez~Perez}(2005)}]{Dorsner:2005fq}%
  \BibitemOpen
  \bibfield  {author} {\bibinfo {author} {\bibfnamefont {I.}~\bibnamefont
  {Dorsner}}\ and\ \bibinfo {author} {\bibfnamefont {P.}~\bibnamefont
  {Fileviez~Perez}},\ }\Doi {10.1016/j.nuclphysb.2005.06.016} {\bibfield
  {journal} {\bibinfo  {journal} {Nucl. Phys. B},\ }\textbf {\bibinfo {volume}
  {723}},\ \bibinfo {pages} {53} (\bibinfo {year} {2005})},\ \Eprint
  {http://arxiv.org/abs/hep-ph/0504276} {arXiv:hep-ph/0504276} \BibitemShut
  {NoStop}%
\bibitem [{\citenamefont {Dorsner}\ \emph {et~al.}(2007)\citenamefont
  {Dorsner}, \citenamefont {Fileviez~Perez},\ and\ \citenamefont
  {Rodrigo}}]{Dorsner:2006hw}%
  \BibitemOpen
  \bibfield  {author} {\bibinfo {author} {\bibfnamefont {I.}~\bibnamefont
  {Dorsner}}, \bibinfo {author} {\bibfnamefont {P.}~\bibnamefont
  {Fileviez~Perez}}, \ and\ \bibinfo {author} {\bibfnamefont {G.}~\bibnamefont
  {Rodrigo}},\ }\Doi {10.1103/PhysRevD.75.125007} {\bibfield  {journal}
  {\bibinfo  {journal} {Phys. Rev. D},\ }\textbf {\bibinfo {volume} {75}},\
  \bibinfo {pages} {125007} (\bibinfo {year} {2007})},\ \Eprint
  {http://arxiv.org/abs/hep-ph/0607208} {arXiv:hep-ph/0607208} \BibitemShut
  {NoStop}%
\bibitem [{\citenamefont {Dorsner}\ and\ \citenamefont
  {Fileviez~Perez}(2007)}]{Dorsner:2006fx}%
  \BibitemOpen
  \bibfield  {author} {\bibinfo {author} {\bibfnamefont {I.}~\bibnamefont
  {Dorsner}}\ and\ \bibinfo {author} {\bibfnamefont {P.}~\bibnamefont
  {Fileviez~Perez}},\ }\Doi {10.1088/1126-6708/2007/06/029} {\bibfield
  {journal} {\bibinfo  {journal} {JHEP},\ }\textbf {\bibinfo {volume} {06}},\
  \bibinfo {pages} {029} (\bibinfo {year} {2007})},\ \Eprint
  {http://arxiv.org/abs/hep-ph/0612216} {arXiv:hep-ph/0612216} \BibitemShut
  {NoStop}%
\bibitem [{\citenamefont {Bajc}\ and\ \citenamefont
  {Senjanovic}(2007)}]{Bajc:2006ia}%
  \BibitemOpen
  \bibfield  {author} {\bibinfo {author} {\bibfnamefont {B.}~\bibnamefont
  {Bajc}}\ and\ \bibinfo {author} {\bibfnamefont {G.}~\bibnamefont
  {Senjanovic}},\ }\Doi {10.1088/1126-6708/2007/08/014} {\bibfield  {journal}
  {\bibinfo  {journal} {JHEP},\ }\textbf {\bibinfo {volume} {08}},\ \bibinfo
  {pages} {014} (\bibinfo {year} {2007})},\ \Eprint
  {http://arxiv.org/abs/hep-ph/0612029} {arXiv:hep-ph/0612029} \BibitemShut
  {NoStop}%
\bibitem [{\citenamefont
  {Fileviez~P\'erez}(2007){\natexlab{a}}}]{Perez:2007rm}%
  \BibitemOpen
  \bibfield  {author} {\bibinfo {author} {\bibfnamefont {P.}~\bibnamefont
  {Fileviez~P\'erez}},\ }\Doi {10.1016/j.physletb.2007.07.075} {\bibfield
  {journal} {\bibinfo  {journal} {Phys. Lett. B},\ }\textbf {\bibinfo {volume}
  {654}},\ \bibinfo {pages} {189} (\bibinfo {year} {2007}{\natexlab{a}})},\
  \Eprint {http://arxiv.org/abs/hep-ph/0702287} {arXiv:hep-ph/0702287}
  \BibitemShut {NoStop}%
\bibitem [{\citenamefont {Bajc}\ \emph
  {et~al.}(2002){\natexlab{a}}\citenamefont {Bajc}, \citenamefont
  {Fileviez~P\'erez},\ and\ \citenamefont {Senjanovic}}]{Bajc:2002bv}%
  \BibitemOpen
  \bibfield  {author} {\bibinfo {author} {\bibfnamefont {B.}~\bibnamefont
  {Bajc}}, \bibinfo {author} {\bibfnamefont {P.}~\bibnamefont
  {Fileviez~P\'erez}}, \ and\ \bibinfo {author} {\bibfnamefont
  {G.}~\bibnamefont {Senjanovic}},\ }\Doi {10.1103/PhysRevD.66.075005}
  {\bibfield  {journal} {\bibinfo  {journal} {Phys. Rev. D},\ }\textbf
  {\bibinfo {volume} {66}},\ \bibinfo {pages} {075005} (\bibinfo {year}
  {2002}{\natexlab{a}})},\ \Eprint {http://arxiv.org/abs/hep-ph/0204311}
  {arXiv:hep-ph/0204311} \BibitemShut {NoStop}%
\bibitem [{\citenamefont {Bajc}\ \emph
  {et~al.}(2002){\natexlab{b}}\citenamefont {Bajc}, \citenamefont
  {Fileviez~P\'erez},\ and\ \citenamefont {Senjanovic}}]{Bajc:2002pg}%
  \BibitemOpen
  \bibfield  {author} {\bibinfo {author} {\bibfnamefont {B.}~\bibnamefont
  {Bajc}}, \bibinfo {author} {\bibfnamefont {P.}~\bibnamefont
  {Fileviez~P\'erez}}, \ and\ \bibinfo {author} {\bibfnamefont
  {G.}~\bibnamefont {Senjanovic}},\ }\href@noop {} { (\bibinfo {year}
  {2002}{\natexlab{b}})},\ \Eprint {http://arxiv.org/abs/hep-ph/0210374}
  {arXiv:hep-ph/0210374} \BibitemShut {NoStop}%
\bibitem [{\citenamefont {Emmanuel-Costa}\ and\ \citenamefont
  {Wiesenfeldt}(2003)}]{EmmanuelCosta:2003pu}%
  \BibitemOpen
  \bibfield  {author} {\bibinfo {author} {\bibfnamefont {D.}~\bibnamefont
  {Emmanuel-Costa}}\ and\ \bibinfo {author} {\bibfnamefont {S.}~\bibnamefont
  {Wiesenfeldt}},\ }\href@noop {} {\bibfield  {journal} {\bibinfo  {journal}
  {Nucl. Phys. B},\ }\textbf {\bibinfo {volume} {661}},\ \bibinfo {pages} {62}
  (\bibinfo {year} {2003})},\ \Eprint {http://arxiv.org/abs/hep-ph/0302272}
  {arXiv:hep-ph/0302272} \BibitemShut {NoStop}%
\bibitem [{\citenamefont
  {Fileviez~P\'erez}(2007){\natexlab{b}}}]{Perez:2007iw}%
  \BibitemOpen
  \bibfield  {author} {\bibinfo {author} {\bibfnamefont {P.}~\bibnamefont
  {Fileviez~P\'erez}},\ }\Doi {10.1103/PhysRevD.76.071701} {\bibfield
  {journal} {\bibinfo  {journal} {Phys.Rev. D},\ }\textbf {\bibinfo {volume}
  {76}},\ \bibinfo {pages} {071701} (\bibinfo {year} {2007}{\natexlab{b}})},\
  \Eprint {http://arxiv.org/abs/0705.3589} {arXiv:0705.3589 [hep-ph]}
  \BibitemShut {NoStop}%
\bibitem [{\citenamefont {Nath}\ and\ \citenamefont
  {Fileviez~P\'erez}(2007)}]{Nath:2006ut}%
  \BibitemOpen
  \bibfield  {author} {\bibinfo {author} {\bibfnamefont {P.}~\bibnamefont
  {Nath}}\ and\ \bibinfo {author} {\bibfnamefont {P.}~\bibnamefont
  {Fileviez~P\'erez}},\ }\Doi {10.1016/j.physrep.2007.02.010} {\bibfield
  {journal} {\bibinfo  {journal} {Phys.Rept.},\ }\textbf {\bibinfo {volume}
  {441}},\ \bibinfo {pages} {191} (\bibinfo {year} {2007})},\ \Eprint
  {http://arxiv.org/abs/hep-ph/0601023} {arXiv:hep-ph/0601023 [hep-ph]}
  \BibitemShut {NoStop}%
\bibitem [{\citenamefont {Branco}\ \emph {et~al.}(1987)\citenamefont {Branco},
  \citenamefont {Geng}, \citenamefont {Marshak},\ and\ \citenamefont
  {Xue}}]{Branco:1987tv}%
  \BibitemOpen
  \bibfield  {author} {\bibinfo {author} {\bibfnamefont {G.~C.}\ \bibnamefont
  {Branco}}, \bibinfo {author} {\bibfnamefont {C.~Q.}\ \bibnamefont {Geng}},
  \bibinfo {author} {\bibfnamefont {R.~E.}\ \bibnamefont {Marshak}}, \ and\
  \bibinfo {author} {\bibfnamefont {P.~Y.}\ \bibnamefont {Xue}},\ }\Doi
  {10.1103/PhysRevD.36.928} {\bibfield  {journal} {\bibinfo  {journal} {Phys.
  Rev. D},\ }\textbf {\bibinfo {volume} {36}},\ \bibinfo {pages} {928}
  (\bibinfo {year} {1987})}\BibitemShut {NoStop}%
\bibitem [{\citenamefont {Babu}\ and\ \citenamefont
  {Kubo}(2005)}]{Babu:2004tn}%
  \BibitemOpen
  \bibfield  {author} {\bibinfo {author} {\bibfnamefont {K.~S.}\ \bibnamefont
  {Babu}}\ and\ \bibinfo {author} {\bibfnamefont {J.}~\bibnamefont {Kubo}},\
  }\Doi {10.1103/PhysRevD.71.056006} {\bibfield  {journal} {\bibinfo  {journal}
  {Phys.Rev. D},\ }\textbf {\bibinfo {volume} {71}},\ \bibinfo {pages} {056006}
  (\bibinfo {year} {2005})},\ \Eprint {http://arxiv.org/abs/hep-ph/0411226}
  {arXiv:hep-ph/0411226 [hep-ph]} \BibitemShut {NoStop}%
\bibitem [{\citenamefont {Grimus}\ \emph {et~al.}(2004)\citenamefont {Grimus},
  \citenamefont {Joshipura}, \citenamefont {Lavoura},\ and\ \citenamefont
  {Tanimoto}}]{Grimus:2004hf}%
  \BibitemOpen
  \bibfield  {author} {\bibinfo {author} {\bibfnamefont {W.}~\bibnamefont
  {Grimus}}, \bibinfo {author} {\bibfnamefont {A.~S.}\ \bibnamefont
  {Joshipura}}, \bibinfo {author} {\bibfnamefont {L.}~\bibnamefont {Lavoura}},
  \ and\ \bibinfo {author} {\bibfnamefont {M.}~\bibnamefont {Tanimoto}},\ }\Doi
  {10.1140/epjc/s2004-01890-5} {\bibfield  {journal} {\bibinfo  {journal} {Eur.
  Phys. J. C},\ }\textbf {\bibinfo {volume} {36}},\ \bibinfo {pages} {227}
  (\bibinfo {year} {2004})},\ \Eprint {http://arxiv.org/abs/hep-ph/0405016}
  {arXiv:hep-ph/0405016} \BibitemShut {NoStop}%
\bibitem [{\citenamefont {Low}(2005)}]{Low:2005yc}%
  \BibitemOpen
  \bibfield  {author} {\bibinfo {author} {\bibfnamefont {C.~I.}\ \bibnamefont
  {Low}},\ }\Doi {10.1103/PhysRevD.71.073007} {\bibfield  {journal} {\bibinfo
  {journal} {Phys. Rev. D},\ }\textbf {\bibinfo {volume} {71}},\ \bibinfo
  {pages} {073007} (\bibinfo {year} {2005})},\ \Eprint
  {http://arxiv.org/abs/hep-ph/0501251} {arXiv:hep-ph/0501251} \BibitemShut
  {NoStop}%
\bibitem [{\citenamefont {Ferreira}\ and\ \citenamefont
  {Silva}(2010)}]{Ferreira:2010ir}%
  \BibitemOpen
  \bibfield  {author} {\bibinfo {author} {\bibfnamefont {P.~M.}\ \bibnamefont
  {Ferreira}}\ and\ \bibinfo {author} {\bibfnamefont {J.~P.}\ \bibnamefont
  {Silva}},\ }\href@noop {} { (\bibinfo {year} {2010})},\ \Eprint
  {http://arxiv.org/abs/1012.2874} {arXiv:1012.2874 [hep-ph]} \BibitemShut
  {NoStop}%
\bibitem [{\citenamefont {Canales}\ and\ \citenamefont
  {Mondragon}(2011)}]{Canales:2011ug}%
  \BibitemOpen
  \bibfield  {author} {\bibinfo {author} {\bibfnamefont {F.~G.}\ \bibnamefont
  {Canales}}\ and\ \bibinfo {author} {\bibfnamefont {A.}~\bibnamefont
  {Mondragon}},\ }\href@noop {} { (\bibinfo {year} {2011})},\ \Eprint
  {http://arxiv.org/abs/1101.3807} {arXiv:1101.3807 [hep-ph]} \BibitemShut
  {NoStop}%
\bibitem [{\citenamefont {Branco}\ \emph {et~al.}(2000)\citenamefont {Branco},
  \citenamefont {Emmanuel-Costa},\ and\ \citenamefont
  {Gonz\'alez~Felipe}}]{Branco:1999nb}%
  \BibitemOpen
  \bibfield  {author} {\bibinfo {author} {\bibfnamefont {G.~C.}\ \bibnamefont
  {Branco}}, \bibinfo {author} {\bibfnamefont {D.}~\bibnamefont
  {Emmanuel-Costa}}, \ and\ \bibinfo {author} {\bibfnamefont {R.}~\bibnamefont
  {Gonz\'alez~Felipe}},\ }\Doi {10.1016/S0370-2693(00)00193-3} {\bibfield
  {journal} {\bibinfo  {journal} {Phys. Lett. B},\ }\textbf {\bibinfo {volume}
  {477}},\ \bibinfo {pages} {147} (\bibinfo {year} {2000})},\ \Eprint
  {http://arxiv.org/abs/hep-ph/9911418} {arXiv:hep-ph/9911418} \BibitemShut
  {NoStop}%
\bibitem [{\citenamefont {Branco}\ \emph {et~al.}(2009)\citenamefont {Branco},
  \citenamefont {Emmanuel-Costa}, \citenamefont {Gonz\'alez~Felipe},\ and\
  \citenamefont {Ser\^odio}}]{Branco:2007nn}%
  \BibitemOpen
  \bibfield  {author} {\bibinfo {author} {\bibfnamefont {G.~C.}\ \bibnamefont
  {Branco}}, \bibinfo {author} {\bibfnamefont {D.}~\bibnamefont
  {Emmanuel-Costa}}, \bibinfo {author} {\bibfnamefont {R.}~\bibnamefont
  {Gonz\'alez~Felipe}}, \ and\ \bibinfo {author} {\bibfnamefont
  {H.}~\bibnamefont {Ser\^odio}},\ }\Doi {10.1016/j.physletb.2008.10.059}
  {\bibfield  {journal} {\bibinfo  {journal} {Phys. Lett. B},\ }\textbf
  {\bibinfo {volume} {670}},\ \bibinfo {pages} {340} (\bibinfo {year}
  {2009})},\ \Eprint {http://arxiv.org/abs/0711.1613} {arXiv:0711.1613
  [hep-ph]} \BibitemShut {NoStop}%
\bibitem [{\citenamefont {Emmanuel-Costa}\ and\ \citenamefont
  {Sim\~oes}(2009)}]{EmmanuelCosta:2009bx}%
  \BibitemOpen
  \bibfield  {author} {\bibinfo {author} {\bibfnamefont {D.}~\bibnamefont
  {Emmanuel-Costa}}\ and\ \bibinfo {author} {\bibfnamefont {C.}~\bibnamefont
  {Sim\~oes}},\ }\Doi {10.1103/PhysRevD.79.073006} {\bibfield  {journal}
  {\bibinfo  {journal} {Phys. Rev. D},\ }\textbf {\bibinfo {volume} {79}},\
  \bibinfo {pages} {073006} (\bibinfo {year} {2009})},\ \Eprint
  {http://arxiv.org/abs/0903.0564} {arXiv:0903.0564 [hep-ph]} \BibitemShut
  {NoStop}%
\bibitem [{\citenamefont {Branco}\ \emph {et~al.}(1989)\citenamefont {Branco},
  \citenamefont {Lavoura},\ and\ \citenamefont {Mota}}]{Branco:1988iq}%
  \BibitemOpen
  \bibfield  {author} {\bibinfo {author} {\bibfnamefont {G.~C.}\ \bibnamefont
  {Branco}}, \bibinfo {author} {\bibfnamefont {L.}~\bibnamefont {Lavoura}}, \
  and\ \bibinfo {author} {\bibfnamefont {F.}~\bibnamefont {Mota}},\ }\Doi
  {10.1103/PhysRevD.39.3443} {\bibfield  {journal} {\bibinfo  {journal} {Phys.
  Rev. D},\ }\textbf {\bibinfo {volume} {39}},\ \bibinfo {pages} {3443}
  (\bibinfo {year} {1989})}\BibitemShut {NoStop}%
\bibitem [{\citenamefont {Fritzsch}(1978)}]{Fritzsch:1977vd}%
  \BibitemOpen
  \bibfield  {author} {\bibinfo {author} {\bibfnamefont {H.}~\bibnamefont
  {Fritzsch}},\ }\Doi {10.1016/0370-2693(78)90524-5} {\bibfield  {journal}
  {\bibinfo  {journal} {Phys. Lett. B},\ }\textbf {\bibinfo {volume} {73}},\
  \bibinfo {pages} {317} (\bibinfo {year} {1978})}\BibitemShut {NoStop}%
\bibitem [{\citenamefont {Li}(1979)}]{Li:1979zj}%
  \BibitemOpen
  \bibfield  {author} {\bibinfo {author} {\bibfnamefont {L.-F.}\ \bibnamefont
  {Li}},\ }\Doi {10.1016/0370-2693(79)91240-1} {\bibfield  {journal} {\bibinfo
  {journal} {Phys. Lett. B},\ }\textbf {\bibinfo {volume} {84}},\ \bibinfo
  {pages} {461} (\bibinfo {year} {1979})}\BibitemShut {NoStop}%
\bibitem [{\citenamefont {Fritzsch}(1979)}]{Fritzsch:1979zq}%
  \BibitemOpen
  \bibfield  {author} {\bibinfo {author} {\bibfnamefont {H.}~\bibnamefont
  {Fritzsch}},\ }\Doi {10.1016/0550-3213(79)90362-6} {\bibfield  {journal}
  {\bibinfo  {journal} {Nucl. Phys. B},\ }\textbf {\bibinfo {volume} {155}},\
  \bibinfo {pages} {189} (\bibinfo {year} {1979})}\BibitemShut {NoStop}%
\bibitem [{\citenamefont {Branco}\ \emph {et~al.}(2010)\citenamefont {Branco},
  \citenamefont {Emmanuel-Costa},\ and\ \citenamefont
  {Sim\~oes}}]{Branco:2010tx}%
  \BibitemOpen
  \bibfield  {author} {\bibinfo {author} {\bibfnamefont {G.}~\bibnamefont
  {Branco}}, \bibinfo {author} {\bibfnamefont {D.}~\bibnamefont
  {Emmanuel-Costa}}, \ and\ \bibinfo {author} {\bibfnamefont {C.}~\bibnamefont
  {Sim\~oes}},\ }\Doi {10.1016/j.physletb.2010.05.009} {\bibfield  {journal}
  {\bibinfo  {journal} {Phys.Lett. B},\ }\textbf {\bibinfo {volume} {690}},\
  \bibinfo {pages} {62} (\bibinfo {year} {2010})},\ \Eprint
  {http://arxiv.org/abs/arXiv:1001.5065} {arXiv:arXiv:1001.5065 [hep-ph]}
  \BibitemShut {NoStop}%
\bibitem [{\citenamefont {Fritzsch}\ \emph {et~al.}(2011)\citenamefont
  {Fritzsch}, \citenamefont {Xing},\ and\ \citenamefont
  {Zhou}}]{Fritzsch:2011cu}%
  \BibitemOpen
  \bibfield  {author} {\bibinfo {author} {\bibfnamefont {H.}~\bibnamefont
  {Fritzsch}}, \bibinfo {author} {\bibfnamefont {Z.-z.}\ \bibnamefont {Xing}},
  \ and\ \bibinfo {author} {\bibfnamefont {Y.-L.}\ \bibnamefont {Zhou}},\
  }\href@noop {} { (\bibinfo {year} {2011})},\ \Eprint
  {http://arxiv.org/abs/1101.4272} {arXiv:1101.4272 [hep-ph]} \BibitemShut
  {NoStop}%
\bibitem [{\citenamefont {Cabibbo}(1963)}]{Cabibbo:1963yz}%
  \BibitemOpen
  \bibfield  {author} {\bibinfo {author} {\bibfnamefont {N.}~\bibnamefont
  {Cabibbo}},\ }\Doi {10.1103/PhysRevLett.10.531} {\bibfield  {journal}
  {\bibinfo  {journal} {Phys. Rev. Lett.},\ }\textbf {\bibinfo {volume} {10}},\
  \bibinfo {pages} {531} (\bibinfo {year} {1963})}\BibitemShut {NoStop}%
\bibitem [{\citenamefont {Kobayashi}\ and\ \citenamefont
  {Maskawa}(1973)}]{Kobayashi:1973fv}%
  \BibitemOpen
  \bibfield  {author} {\bibinfo {author} {\bibfnamefont {M.}~\bibnamefont
  {Kobayashi}}\ and\ \bibinfo {author} {\bibfnamefont {T.}~\bibnamefont
  {Maskawa}},\ }\Doi {10.1143/PTP.49.652} {\bibfield  {journal} {\bibinfo
  {journal} {Prog. Theor. Phys.},\ }\textbf {\bibinfo {volume} {49}},\ \bibinfo
  {pages} {652} (\bibinfo {year} {1973})}\BibitemShut {NoStop}%
\bibitem [{\citenamefont {Minkowski}(1977)}]{Minkowski:1977sc}%
  \BibitemOpen
  \bibfield  {author} {\bibinfo {author} {\bibfnamefont {P.}~\bibnamefont
  {Minkowski}},\ }\Doi {10.1016/0370-2693(77)90435-X} {\bibfield  {journal}
  {\bibinfo  {journal} {Phys.Lett. B},\ }\textbf {\bibinfo {volume} {67}},\
  \bibinfo {pages} {421} (\bibinfo {year} {1977})}\BibitemShut {NoStop}%
\bibitem [{\citenamefont {Yanagida}()}]{Yanagida:1979as}%
  \BibitemOpen
  \bibfield  {author} {\bibinfo {author} {\bibfnamefont {T.}~\bibnamefont
  {Yanagida}},\ }\href@noop {} {}\bibinfo {note} {In Proc. of the Workshop on
  Unified Theory and Baryon Number in the Universe, KEK, March
  1979}\BibitemShut {NoStop}%
\bibitem [{\citenamefont {Gell-Mann}\ \emph {et~al.}()\citenamefont
  {Gell-Mann}, \citenamefont {Ramond},\ and\ \citenamefont
  {Slansky}}]{GellMann:1980vs}%
  \BibitemOpen
  \bibfield  {author} {\bibinfo {author} {\bibfnamefont {M.}~\bibnamefont
  {Gell-Mann}}, \bibinfo {author} {\bibfnamefont {P.}~\bibnamefont {Ramond}}, \
  and\ \bibinfo {author} {\bibfnamefont {R.}~\bibnamefont {Slansky}},\
  }\href@noop {} {}\bibinfo {note} {To be published in Supergravity, P. van
  Nieuwenhuizen $\&$ D.Z. Freedman (eds.), North Holland Publ. Co.,
  1979}\BibitemShut {NoStop}%
\bibitem [{\citenamefont {Mohapatra}\ and\ \citenamefont
  {Senjanovic}(1980)}]{Mohapatra:1979ia}%
  \BibitemOpen
  \bibfield  {author} {\bibinfo {author} {\bibfnamefont {R.~N.}\ \bibnamefont
  {Mohapatra}}\ and\ \bibinfo {author} {\bibfnamefont {G.}~\bibnamefont
  {Senjanovic}},\ }\Doi {10.1103/PhysRevLett.44.912} {\bibfield  {journal}
  {\bibinfo  {journal} {Phys.Rev.Lett.},\ }\textbf {\bibinfo {volume} {44}},\
  \bibinfo {pages} {912} (\bibinfo {year} {1980})}\BibitemShut {NoStop}%
\bibitem [{\citenamefont {Li}(1974)}]{Li:1973mq}%
  \BibitemOpen
  \bibfield  {author} {\bibinfo {author} {\bibfnamefont {L.-F.}\ \bibnamefont
  {Li}},\ }\Doi {10.1103/PhysRevD.9.1723} {\bibfield  {journal} {\bibinfo
  {journal} {Phys. Rev. D},\ }\textbf {\bibinfo {volume} {9}},\ \bibinfo
  {pages} {1723} (\bibinfo {year} {1974})}\BibitemShut {NoStop}%
\bibitem [{\citenamefont {Buccella}\ \emph {et~al.}(1980)\citenamefont
  {Buccella}, \citenamefont {Ruegg},\ and\ \citenamefont
  {Savoy}}]{Buccella:1979sk}%
  \BibitemOpen
  \bibfield  {author} {\bibinfo {author} {\bibfnamefont {F.}~\bibnamefont
  {Buccella}}, \bibinfo {author} {\bibfnamefont {H.}~\bibnamefont {Ruegg}}, \
  and\ \bibinfo {author} {\bibfnamefont {C.~A.}\ \bibnamefont {Savoy}},\ }\Doi
  {10.1016/0550-3213(80)90252-7} {\bibfield  {journal} {\bibinfo  {journal}
  {Nucl. Phys. B},\ }\textbf {\bibinfo {volume} {169}},\ \bibinfo {pages} {68}
  (\bibinfo {year} {1980})}\BibitemShut {NoStop}%
\bibitem [{\citenamefont {Guth}\ and\ \citenamefont {Tye}(1980)}]{Guth:1979bh}%
  \BibitemOpen
  \bibfield  {author} {\bibinfo {author} {\bibfnamefont {A.~H.}\ \bibnamefont
  {Guth}}\ and\ \bibinfo {author} {\bibfnamefont {S.~H.~H.}\ \bibnamefont
  {Tye}},\ }\Doi {10.1103/PhysRevLett.44.631} {\bibfield  {journal} {\bibinfo
  {journal} {Phys. Rev. Lett.},\ }\textbf {\bibinfo {volume} {44}},\ \bibinfo
  {pages} {631} (\bibinfo {year} {1980})},\ \bibinfo {note} {[Erratum-ibid.\
  {\bf 44} (1980) 963]}\BibitemShut {NoStop}%
\bibitem [{\citenamefont {Ruegg}(1980)}]{Ruegg:1980gf}%
  \BibitemOpen
  \bibfield  {author} {\bibinfo {author} {\bibfnamefont {H.}~\bibnamefont
  {Ruegg}},\ }\Doi {10.1103/PhysRevD.22.2040} {\bibfield  {journal} {\bibinfo
  {journal} {Phys. Rev. D},\ }\textbf {\bibinfo {volume} {22}},\ \bibinfo
  {pages} {2040} (\bibinfo {year} {1980})}\BibitemShut {NoStop}%
\bibitem [{\citenamefont {Georgi}\ and\ \citenamefont
  {Pais}(1974)}]{Georgi:1974au}%
  \BibitemOpen
  \bibfield  {author} {\bibinfo {author} {\bibfnamefont {H.}~\bibnamefont
  {Georgi}}\ and\ \bibinfo {author} {\bibfnamefont {A.}~\bibnamefont {Pais}},\
  }\Doi {10.1103/PhysRevD.10.1246} {\bibfield  {journal} {\bibinfo  {journal}
  {Phys. Rev. D},\ }\textbf {\bibinfo {volume} {10}},\ \bibinfo {pages} {1246}
  (\bibinfo {year} {1974})}\BibitemShut {NoStop}%
\bibitem [{\citenamefont {Weinberg}(1972)}]{Weinberg:1972fn}%
  \BibitemOpen
  \bibfield  {author} {\bibinfo {author} {\bibfnamefont {S.}~\bibnamefont
  {Weinberg}},\ }\Doi {10.1103/PhysRevLett.29.1698} {\bibfield  {journal}
  {\bibinfo  {journal} {Phys. Rev. Lett.},\ }\textbf {\bibinfo {volume} {29}},\
  \bibinfo {pages} {1698} (\bibinfo {year} {1972})}\BibitemShut {NoStop}%
\bibitem [{\citenamefont {Buras}\ \emph {et~al.}(1978)\citenamefont {Buras},
  \citenamefont {Ellis}, \citenamefont {Gaillard},\ and\ \citenamefont
  {Nanopoulos}}]{Buras:1977yy}%
  \BibitemOpen
  \bibfield  {author} {\bibinfo {author} {\bibfnamefont {A.~J.}\ \bibnamefont
  {Buras}}, \bibinfo {author} {\bibfnamefont {J.~R.}\ \bibnamefont {Ellis}},
  \bibinfo {author} {\bibfnamefont {M.~K.}\ \bibnamefont {Gaillard}}, \ and\
  \bibinfo {author} {\bibfnamefont {D.~V.}\ \bibnamefont {Nanopoulos}},\ }\Doi
  {10.1016/0550-3213(78)90214-6} {\bibfield  {journal} {\bibinfo  {journal}
  {Nucl. Phys. B},\ }\textbf {\bibinfo {volume} {135}},\ \bibinfo {pages} {66}
  (\bibinfo {year} {1978})}\BibitemShut {NoStop}%
\bibitem [{\citenamefont {Nakamura}\ \emph {et~al.}(2010)\citenamefont
  {Nakamura} \emph {et~al.}}]{Nakamura:2010zzi}%
  \BibitemOpen
  \bibfield  {author} {\bibinfo {author} {\bibfnamefont {K.}~\bibnamefont
  {Nakamura}} \emph {et~al.} (\bibinfo {collaboration} {Particle Data Group}),\
  }\Doi {10.1088/0954-3899/37/7A/075021} {\bibfield  {journal} {\bibinfo
  {journal} {J. Phys. G},\ }\textbf {\bibinfo {volume} {37}},\ \bibinfo {pages}
  {075021} (\bibinfo {year} {2010})}\BibitemShut {NoStop}%
\bibitem [{\citenamefont {Langacker}(1981)}]{Langacker:1980js}%
  \BibitemOpen
  \bibfield  {author} {\bibinfo {author} {\bibfnamefont {P.}~\bibnamefont
  {Langacker}},\ }\Doi {10.1016/0370-1573(81)90059-4} {\bibfield  {journal}
  {\bibinfo  {journal} {Phys. Rept.},\ }\textbf {\bibinfo {volume} {72}},\
  \bibinfo {pages} {185} (\bibinfo {year} {1981})}\BibitemShut {NoStop}%
\bibitem [{\citenamefont {Pontecorvo}(1957)}]{Pontecorvo:1957cp}%
  \BibitemOpen
  \bibfield  {author} {\bibinfo {author} {\bibfnamefont {B.}~\bibnamefont
  {Pontecorvo}},\ }\href@noop {} {\bibfield  {journal} {\bibinfo  {journal}
  {Sov. Phys. JETP},\ }\textbf {\bibinfo {volume} {6}},\ \bibinfo {pages} {429}
  (\bibinfo {year} {1957})}\BibitemShut {NoStop}%
\bibitem [{\citenamefont {Pontecorvo}(1958)}]{Pontecorvo:1957qd}%
  \BibitemOpen
  \bibfield  {author} {\bibinfo {author} {\bibfnamefont {B.}~\bibnamefont
  {Pontecorvo}},\ }\href@noop {} {\bibfield  {journal} {\bibinfo  {journal}
  {Sov. Phys. JETP},\ }\textbf {\bibinfo {volume} {7}},\ \bibinfo {pages} {172}
  (\bibinfo {year} {1958})}\BibitemShut {NoStop}%
\bibitem [{\citenamefont {Maki}\ \emph {et~al.}(1962)\citenamefont {Maki},
  \citenamefont {Nakagawa},\ and\ \citenamefont {Sakata}}]{Maki:1962mu}%
  \BibitemOpen
  \bibfield  {author} {\bibinfo {author} {\bibfnamefont {Z.}~\bibnamefont
  {Maki}}, \bibinfo {author} {\bibfnamefont {M.}~\bibnamefont {Nakagawa}}, \
  and\ \bibinfo {author} {\bibfnamefont {S.}~\bibnamefont {Sakata}},\ }\Doi
  {10.1143/PTP.28.870} {\bibfield  {journal} {\bibinfo  {journal} {Prog. Theor.
  Phys.},\ }\textbf {\bibinfo {volume} {28}},\ \bibinfo {pages} {870} (\bibinfo
  {year} {1962})}\BibitemShut {NoStop}%
\bibitem [{\citenamefont {Schwetz}\ \emph {et~al.}(2008)\citenamefont
  {Schwetz}, \citenamefont {Tortola},\ and\ \citenamefont
  {Valle}}]{Schwetz:2008er}%
  \BibitemOpen
  \bibfield  {author} {\bibinfo {author} {\bibfnamefont {T.}~\bibnamefont
  {Schwetz}}, \bibinfo {author} {\bibfnamefont {M.}~\bibnamefont {Tortola}}, \
  and\ \bibinfo {author} {\bibfnamefont {J.~W.}\ \bibnamefont {Valle}},\ }\Doi
  {10.1088/1367-2630/10/11/113011} {\bibfield  {journal} {\bibinfo  {journal}
  {New J.Phys.},\ }\textbf {\bibinfo {volume} {10}},\ \bibinfo {pages} {113011}
  (\bibinfo {year} {2008})},\ \Eprint {http://arxiv.org/abs/arXiv:0808.2016}
  {arXiv:arXiv:0808.2016 [hep-ph]} \BibitemShut {NoStop}%
\bibitem [{\citenamefont {Pascoli}\ \emph {et~al.}(2002)\citenamefont
  {Pascoli}, \citenamefont {Petcov},\ and\ \citenamefont
  {Wolfenstein}}]{Pascoli:2001by}%
  \BibitemOpen
  \bibfield  {author} {\bibinfo {author} {\bibfnamefont {S.}~\bibnamefont
  {Pascoli}}, \bibinfo {author} {\bibfnamefont {S.~T.}\ \bibnamefont {Petcov}},
  \ and\ \bibinfo {author} {\bibfnamefont {L.}~\bibnamefont {Wolfenstein}},\
  }\Doi {10.1016/S0370-2693(01)01403-4} {\bibfield  {journal} {\bibinfo
  {journal} {Phys. Lett. B},\ }\textbf {\bibinfo {volume} {524}},\ \bibinfo
  {pages} {319} (\bibinfo {year} {2002})},\ \Eprint
  {http://arxiv.org/abs/hep-ph/0110287} {arXiv:hep-ph/0110287} \BibitemShut
  {NoStop}%
\bibitem [{\citenamefont {Pascoli}\ and\ \citenamefont
  {Petcov}(2002)}]{Pascoli:2002xq}%
  \BibitemOpen
  \bibfield  {author} {\bibinfo {author} {\bibfnamefont {S.}~\bibnamefont
  {Pascoli}}\ and\ \bibinfo {author} {\bibfnamefont {S.~T.}\ \bibnamefont
  {Petcov}},\ }\Doi {10.1016/S0370-2693(02)02510-8} {\bibfield  {journal}
  {\bibinfo  {journal} {Phys. Lett. B},\ }\textbf {\bibinfo {volume} {544}},\
  \bibinfo {pages} {239} (\bibinfo {year} {2002})},\ \Eprint
  {http://arxiv.org/abs/hep-ph/0205022} {arXiv:hep-ph/0205022} \BibitemShut
  {NoStop}%
\bibitem [{\citenamefont {Pascoli}\ and\ \citenamefont
  {Petcov}(2004)}]{Pascoli:2003ke}%
  \BibitemOpen
  \bibfield  {author} {\bibinfo {author} {\bibfnamefont {S.}~\bibnamefont
  {Pascoli}}\ and\ \bibinfo {author} {\bibfnamefont {S.~T.}\ \bibnamefont
  {Petcov}},\ }\Doi {10.1016/j.physletb.2003.11.030} {\bibfield  {journal}
  {\bibinfo  {journal} {Phys. Lett. B},\ }\textbf {\bibinfo {volume} {580}},\
  \bibinfo {pages} {280} (\bibinfo {year} {2004})},\ \Eprint
  {http://arxiv.org/abs/hep-ph/0310003} {arXiv:hep-ph/0310003} \BibitemShut
  {NoStop}%
\bibitem [{\citenamefont {Spergel}\ \emph {et~al.}(2007)\citenamefont {Spergel}
  \emph {et~al.}}]{Spergel:2006hy}%
  \BibitemOpen
  \bibfield  {author} {\bibinfo {author} {\bibfnamefont {D.~N.}\ \bibnamefont
  {Spergel}} \emph {et~al.} (\bibinfo {collaboration} {WMAP}),\ }\Doi
  {10.1086/513700} {\bibfield  {journal} {\bibinfo  {journal} {Astrophys. J.
  Suppl.},\ }\textbf {\bibinfo {volume} {170}},\ \bibinfo {pages} {377}
  (\bibinfo {year} {2007})},\ \Eprint {http://arxiv.org/abs/astro-ph/0603449}
  {arXiv:astro-ph/0603449} \BibitemShut {NoStop}%
\bibitem [{\citenamefont {Bilenky}\ \emph
  {et~al.}(2001){\natexlab{a}}\citenamefont {Bilenky}, \citenamefont
  {Pascoli},\ and\ \citenamefont {Petcov}}]{Bilenky:2001rz}%
  \BibitemOpen
  \bibfield  {author} {\bibinfo {author} {\bibfnamefont {S.~M.}\ \bibnamefont
  {Bilenky}}, \bibinfo {author} {\bibfnamefont {S.}~\bibnamefont {Pascoli}}, \
  and\ \bibinfo {author} {\bibfnamefont {S.~T.}\ \bibnamefont {Petcov}},\ }\Doi
  {10.1103/PhysRevD.64.053010} {\bibfield  {journal} {\bibinfo  {journal}
  {Phys. Rev. D},\ }\textbf {\bibinfo {volume} {64}},\ \bibinfo {pages}
  {053010} (\bibinfo {year} {2001}{\natexlab{a}})},\ \Eprint
  {http://arxiv.org/abs/hep-ph/0102265} {arXiv:hep-ph/0102265} \BibitemShut
  {NoStop}%
\bibitem [{\citenamefont {Bilenky}\ \emph
  {et~al.}(2001){\natexlab{b}}\citenamefont {Bilenky}, \citenamefont
  {Pascoli},\ and\ \citenamefont {Petcov}}]{Bilenky:2001xq}%
  \BibitemOpen
  \bibfield  {author} {\bibinfo {author} {\bibfnamefont {S.~M.}\ \bibnamefont
  {Bilenky}}, \bibinfo {author} {\bibfnamefont {S.}~\bibnamefont {Pascoli}}, \
  and\ \bibinfo {author} {\bibfnamefont {S.~T.}\ \bibnamefont {Petcov}},\ }\Doi
  {10.1103/PhysRevD.64.113003} {\bibfield  {journal} {\bibinfo  {journal}
  {Phys. Rev. D},\ }\textbf {\bibinfo {volume} {64}},\ \bibinfo {pages}
  {113003} (\bibinfo {year} {2001}{\natexlab{b}})},\ \Eprint
  {http://arxiv.org/abs/hep-ph/0104218} {arXiv:hep-ph/0104218} \BibitemShut
  {NoStop}%
\bibitem [{\citenamefont {Petcov}(2005)}]{Petcov:2005yq}%
  \BibitemOpen
  \bibfield  {author} {\bibinfo {author} {\bibfnamefont {S.~T.}\ \bibnamefont
  {Petcov}},\ }\Doi {10.1088/0031-8949/2005/T121/013} {\bibfield  {journal}
  {\bibinfo  {journal} {Phys. Scripta},\ }\textbf {\bibinfo {volume} {T121}},\
  \bibinfo {pages} {94} (\bibinfo {year} {2005})},\ \Eprint
  {http://arxiv.org/abs/hep-ph/0504166} {arXiv:hep-ph/0504166} \BibitemShut
  {NoStop}%
\bibitem [{\citenamefont {Thomas}\ \emph {et~al.}(2010)\citenamefont {Thomas},
  \citenamefont {Abdalla},\ and\ \citenamefont {Lahav}}]{Thomas:2009ae}%
  \BibitemOpen
  \bibfield  {author} {\bibinfo {author} {\bibfnamefont {S.~A.}\ \bibnamefont
  {Thomas}}, \bibinfo {author} {\bibfnamefont {F.~B.}\ \bibnamefont {Abdalla}},
  \ and\ \bibinfo {author} {\bibfnamefont {O.}~\bibnamefont {Lahav}},\ }\Doi
  {10.1103/PhysRevLett.105.031301} {\bibfield  {journal} {\bibinfo  {journal}
  {Phys. Rev. Lett.},\ }\textbf {\bibinfo {volume} {105}},\ \bibinfo {pages}
  {031301} (\bibinfo {year} {2010})},\ \Eprint {http://arxiv.org/abs/0911.5291}
  {arXiv:0911.5291 [astro-ph.CO]} \BibitemShut {NoStop}%
\bibitem [{\citenamefont {Fusaoka}\ and\ \citenamefont
  {Koide}(1998)}]{Fusaoka:1998vc}%
  \BibitemOpen
  \bibfield  {author} {\bibinfo {author} {\bibfnamefont {H.}~\bibnamefont
  {Fusaoka}}\ and\ \bibinfo {author} {\bibfnamefont {Y.}~\bibnamefont
  {Koide}},\ }\Doi {10.1103/PhysRevD.57.3986} {\bibfield  {journal} {\bibinfo
  {journal} {Phys.Rev. D},\ }\textbf {\bibinfo {volume} {57}},\ \bibinfo
  {pages} {3986} (\bibinfo {year} {1998})},\ \Eprint
  {http://arxiv.org/abs/hep-ph/9712201} {arXiv:hep-ph/9712201 [hep-ph]}
  \BibitemShut {NoStop}%
\bibitem [{\citenamefont {Xing}\ \emph {et~al.}(2008)\citenamefont {Xing},
  \citenamefont {Zhang},\ and\ \citenamefont {Zhou}}]{Xing:2007fb}%
  \BibitemOpen
  \bibfield  {author} {\bibinfo {author} {\bibfnamefont {Z.-z.}\ \bibnamefont
  {Xing}}, \bibinfo {author} {\bibfnamefont {H.}~\bibnamefont {Zhang}}, \ and\
  \bibinfo {author} {\bibfnamefont {S.}~\bibnamefont {Zhou}},\ }\Doi
  {10.1103/PhysRevD.77.113016} {\bibfield  {journal} {\bibinfo  {journal}
  {Phys.Rev. D},\ }\textbf {\bibinfo {volume} {77}},\ \bibinfo {pages} {113016}
  (\bibinfo {year} {2008})},\ \Eprint {http://arxiv.org/abs/0712.1419}
  {arXiv:0712.1419 [hep-ph]} \BibitemShut {NoStop}%
\bibitem [{\citenamefont {Branco}\ and\ \citenamefont
  {Mota}(1992)}]{Branco:1992ba}%
  \BibitemOpen
  \bibfield  {author} {\bibinfo {author} {\bibfnamefont {G.~C.}\ \bibnamefont
  {Branco}}\ and\ \bibinfo {author} {\bibfnamefont {F.}~\bibnamefont {Mota}},\
  }\Doi {10.1016/0370-2693(92)90781-X} {\bibfield  {journal} {\bibinfo
  {journal} {Phys. Lett. B},\ }\textbf {\bibinfo {volume} {280}},\ \bibinfo
  {pages} {109} (\bibinfo {year} {1992})}\BibitemShut {NoStop}%
\bibitem [{\citenamefont {Fogli}\ \emph {et~al.}(2008)\citenamefont {Fogli},
  \citenamefont {Lisi}, \citenamefont {Marrone}, \citenamefont {Palazzo},\ and\
  \citenamefont {Rotunno}}]{Fogli:2008jx}%
  \BibitemOpen
  \bibfield  {author} {\bibinfo {author} {\bibfnamefont {G.~L.}\ \bibnamefont
  {Fogli}}, \bibinfo {author} {\bibfnamefont {E.}~\bibnamefont {Lisi}},
  \bibinfo {author} {\bibfnamefont {A.}~\bibnamefont {Marrone}}, \bibinfo
  {author} {\bibfnamefont {A.}~\bibnamefont {Palazzo}}, \ and\ \bibinfo
  {author} {\bibfnamefont {A.~M.}\ \bibnamefont {Rotunno}},\ }\Doi
  {10.1103/PhysRevLett.101.141801} {\bibfield  {journal} {\bibinfo  {journal}
  {Phys. Rev. Lett.},\ }\textbf {\bibinfo {volume} {101}},\ \bibinfo {pages}
  {141801} (\bibinfo {year} {2008})},\ \Eprint {http://arxiv.org/abs/0806.2649}
  {arXiv:0806.2649 [hep-ph]} \BibitemShut {NoStop}%
\bibitem [{\citenamefont {Rodrigo}\ and\ \citenamefont
  {Santamaria}(1993)}]{Rodrigo:1993hc}%
  \BibitemOpen
  \bibfield  {author} {\bibinfo {author} {\bibfnamefont {G.}~\bibnamefont
  {Rodrigo}}\ and\ \bibinfo {author} {\bibfnamefont {A.}~\bibnamefont
  {Santamaria}},\ }\Doi {10.1016/0370-2693(93)90016-B} {\bibfield  {journal}
  {\bibinfo  {journal} {Phys.Lett. B},\ }\textbf {\bibinfo {volume} {313}},\
  \bibinfo {pages} {441} (\bibinfo {year} {1993})},\ \Eprint
  {http://arxiv.org/abs/hep-ph/9305305} {arXiv:hep-ph/9305305 [hep-ph]}
  \BibitemShut {NoStop}%
\bibitem [{\citenamefont {Chetyrkin}\ \emph {et~al.}(1996)\citenamefont
  {Chetyrkin}, \citenamefont {Kuhn},\ and\ \citenamefont
  {Kwiatkowski}}]{Chetyrkin:1996ia}%
  \BibitemOpen
  \bibfield  {author} {\bibinfo {author} {\bibfnamefont {K.}~\bibnamefont
  {Chetyrkin}}, \bibinfo {author} {\bibfnamefont {J.~H.}\ \bibnamefont {Kuhn}},
  \ and\ \bibinfo {author} {\bibfnamefont {A.}~\bibnamefont {Kwiatkowski}},\
  }\Doi {10.1016/S0370-1573(96)00012-9} {\bibfield  {journal} {\bibinfo
  {journal} {Phys.Rept.},\ }\textbf {\bibinfo {volume} {277}},\ \bibinfo
  {pages} {189} (\bibinfo {year} {1996})},\ \bibinfo {note} {revised
  version}\BibitemShut {NoStop}%
\bibitem [{\citenamefont {van Ritbergen}\ \emph {et~al.}(1997)\citenamefont
  {van Ritbergen}, \citenamefont {Vermaseren},\ and\ \citenamefont
  {Larin}}]{vanRitbergen:1997va}%
  \BibitemOpen
  \bibfield  {author} {\bibinfo {author} {\bibfnamefont {T.}~\bibnamefont {van
  Ritbergen}}, \bibinfo {author} {\bibfnamefont {J.}~\bibnamefont
  {Vermaseren}}, \ and\ \bibinfo {author} {\bibfnamefont {S.}~\bibnamefont
  {Larin}},\ }\Doi {10.1016/S0370-2693(97)00370-5} {\bibfield  {journal}
  {\bibinfo  {journal} {Phys.Lett. B},\ }\textbf {\bibinfo {volume} {400}},\
  \bibinfo {pages} {379} (\bibinfo {year} {1997})},\ \Eprint
  {http://arxiv.org/abs/hep-ph/9701390} {arXiv:hep-ph/9701390 [hep-ph]}
  \BibitemShut {NoStop}%
\bibitem [{\citenamefont {Jones}(1982)}]{Jones:1981we}%
  \BibitemOpen
  \bibfield  {author} {\bibinfo {author} {\bibfnamefont {D.~R.~T.}\
  \bibnamefont {Jones}},\ }\Doi {10.1103/PhysRevD.25.581} {\bibfield  {journal}
  {\bibinfo  {journal} {Phys. Rev. D},\ }\textbf {\bibinfo {volume} {25}},\
  \bibinfo {pages} {581} (\bibinfo {year} {1982})}\BibitemShut {NoStop}%
\bibitem [{\citenamefont {Machacek}\ and\ \citenamefont
  {Vaughn}(1983)}]{Machacek:1983tz}%
  \BibitemOpen
  \bibfield  {author} {\bibinfo {author} {\bibfnamefont {M.~E.}\ \bibnamefont
  {Machacek}}\ and\ \bibinfo {author} {\bibfnamefont {M.~T.}\ \bibnamefont
  {Vaughn}},\ }\Doi {10.1016/0550-3213(83)90610-7} {\bibfield  {journal}
  {\bibinfo  {journal} {Nucl. Phys. B},\ }\textbf {\bibinfo {volume} {222}},\
  \bibinfo {pages} {83} (\bibinfo {year} {1983})}\BibitemShut {NoStop}%
\bibitem [{\citenamefont {Luo}\ \emph {et~al.}(2003)\citenamefont {Luo},
  \citenamefont {Wang},\ and\ \citenamefont {Xiao}}]{Luo:2002ti}%
  \BibitemOpen
  \bibfield  {author} {\bibinfo {author} {\bibfnamefont {M.-x.}\ \bibnamefont
  {Luo}}, \bibinfo {author} {\bibfnamefont {H.-w.}\ \bibnamefont {Wang}}, \
  and\ \bibinfo {author} {\bibfnamefont {Y.}~\bibnamefont {Xiao}},\ }\Doi
  {10.1103/PhysRevD.67.065019} {\bibfield  {journal} {\bibinfo  {journal}
  {Phys. Rev. D},\ }\textbf {\bibinfo {volume} {67}},\ \bibinfo {pages}
  {065019} (\bibinfo {year} {2003})},\ \Eprint
  {http://arxiv.org/abs/hep-ph/0211440} {arXiv:hep-ph/0211440} \BibitemShut
  {NoStop}%
\end{thebibliography}%

\end{document}